\documentstyle[a4p,12pt,cite,epsfig]{article}
\begin{document}
\renewcommand{\arraystretch}{1.0} 
\newcommand{\dfrac}{\frac}
\begin{titlepage}
\begin{flushright}
%TAUP-2872-08
\end{flushright}
\begin{center}
\vspace{1cm}

{\LARGE\it  Quark-Antiquark Energy Density Function applied

\vspace{2mm}            

to Di-Gauge Boson Production at the LHC} 

\vspace*{1.9mm}

\vspace{0.9cm}
{\large G}{\small IDEON} {\large A}{\small LEXANDER}\footnote{ 
e-mail: alex@atlas2.tau.ac.il} and 
{\large E}{\small REZ} {\large R}{\small EINHERZ-}{\large A}{\small RONIS}\footnote
{e-mail: erezra@atlas2.tau.ac.il}\\ 
\vspace{0.4cm}

{\large\it Raymond and Beverly Sackler School of Physics and Astronomy\\
Tel-Aviv University,
Tel-Aviv 69978, Israel}\\
\end{center}
\vspace{0.9cm}

\begin{abstract}
\noindent
{\small In view of the start up of the 14 TeV $pp$ Large Hadron
Collider 
the quark anti$-$quark 
reactions leading to the 
final states
$W^+W^-$, $W^{\pm}Z^0$ and $Z^0Z^0$ are studied, in the frame work
of the 
Standard Model $(SM)$, using
helicity amplitudes. The differential and 
total cross sections are first evaluated 
in the parton anti-parton center of mass system. Subsequently they are 
transformed to their expected structure
in $pp$ collisions through the parton anti-parton Energy Density Functions 
which are here derived from
the known Parton Density Functions of the proton.  
In particular the single and joint longitudinal
polarizations of the di-boson final states are calculated. 
The effect on these reactions from the presence of s-channel  
heavy vector bosons, like the $W'$ and $Z'$, 
are evaluated to explore
the possibility to utilize the gauge boson pair production 
as a probe for these 'Beyond the $SM$' phenomena.} 
\end{abstract}
\begin{flushleft}
\hspace*{+1.05cm}PACS numbers: 13.85.t, 13.88.+e, 14.70.-e\\
\end{flushleft}
\vspace{2cm}
%\centering{(\today)}
\vspace{2mm}
\end{titlepage}
\section{Introduction}

In high energy proton-proton ($pp$) colliders, such as the 
LHC (Large Hadron Collider) at CERN with a center of mass energy 
$\sqrt{S}$=14 TeV,
the production of heavy
gauge vector boson pairs ($VV'$) occurs
predominantly by the quark anti$-$quark 
($q {\bar q}$) reactions
and to a lesser extent via processes involving gluons. 
In the planned International
Linear Collider (ILC) these pairs of bosons will be produced in the 
$e^+ e^-$ colliding beams at their center of mass energy of
500 GeV up to $\simeq$1000 GeV \cite{ilc}
which practically
coincides with the center of mass energy supplied by the
accelerator. 
This is not the case in the LHC 
where the center of mass energy of the
parton anti-parton system, here denoted by $Ecm$, varies over a wide range, 
essentially from zero up to 14 TeV. This feature, 
that for some applications
is a drawback, has the advantage that it offers the possibility
to cover a very wide $Ecm$ region over which
the search of new particle states can be conducted.\\

Some properties of the total and
differential cross sections such as the transverse momenta of the
$W^+ W^-$, $W^{\pm} Z^0$ and $Z^0 Z^0$ final states 
produced via $q \bar q$ reactions,  
have been estimated \cite{nuss} via the helicity amplitude
technique. This method which was adopted in this work, was 
applied to $pp$ and $e^+e^-$ 
collisions, in order to extend the former studies
and to estimate the  
Standard Model ($SM$) expectations
for cross sections including single and joint  
longitudinal polarizations of the $VV'$ boson pairs. 
The total and differential cross sections, and in
particular the evaluation of longitudinal polarizations, are
examined with the aim to asses their power to detect  
'beyond the $SM$' phenomena like the existence of heavy $Z$
vector bosons, referred to as $Z^{'}$, and   
similar states expected from the conjecture of extra dimensions 
$(ED)$ set of $Z^{\star}$ states. 
Our findings
can directly be applied to results of future high energy
$e^+e^-$ colliders like the ILC and/or 
the Compact Linear Collider (CLIC).
The comparison of our calculations with some of the properties
envisaged in $pp$ reactions at the LHC, like cross sections, 
requires the incorporation of
the effects arising from the 
Parton Density Function ($PDF$) on the quark anti$-$quark
center of mass energy distributions as well as  
a proper handling of the contributions arising 
from processes involving gluons.\\

In the following 
Section \ref{sec_pdf} we describe our method to obtain
an expression for 
the parton anti-parton $Ecm$ distributions in $pp$ collisions,
here denoted by $EDF$. 
In Section \ref{amplitudes} a 
brief outline of our helicity amplitude calculations for cross
sections and polarizations are presented. 
Section \ref{sec_ww} is devoted to 
the $SM$ reactions of 
$q \bar q$, $q \bar q'$ and $e^+e^-$ leading to the
final states $W^+ W^-$, $W^{\pm}Z^0$ and  $Z^0Z^0$ 
and their manifestation in $pp$ collisions using the $EDF$.
In the same Section we also test our $EDF$ based method
to transform $q \bar q$ processes to $pp$ reactions 
by comparing it to a Pythia Monte Carlo generated 
$pp \to Z^0Z^0$ sample. 
In Section \ref{massive} we examine the effect of
the presence of massive $Z'$ and $W'$ bosons in the
s$-$channel 
on the cross sections and 
longitudinal polarizations of the $W^+W^-$ and $W^-Z^0$
final states. Finally we conclude in Section \ref{summary}
with a short summary.
\section{Parameterization of the Energy Density Functions}
\label{sec_pdf}
\subsection{The need for the Energy Density Functions}
In general, theoretical calculations of cross sections
and other properties of parton anti-parton  
reactions
leading to exclusive final states, such as
$\bar q q \to W^+ W^-$, are carried out as a function of the
parton anti-parton center of mass energy, $Ecm$. In the study of
the proton-proton collisions, at a given
center of mass energy $\sqrt{S_{pp}}$, leading to the identical
final state 
the $Ecm$ can be determined event by event from the momenta of this final 
state particles. The 
question however is how the theoretical calculated cross section 
dependence on $Ecm$ , e.g. 
$\sigma(\bar q q\to W^+ W^-)$, is transformed when measured in
$pp$ collisions.
In the interacting $pp$ system  
there exist
an infinite continuous set of colliding parton anti-parton pairs 
which result in the very same $Ecm$ value. The reason
stems from the fact that the partons of the protons have
a continuous energy distribution, given by the Parton Density Function, 
$PDF$, which describe the relative parton energy  
with respect to that of the proton laboratory energy.
Thus the multi occurrence of parton-pair $Ecm$ values in $pp$ collisions
results in a non-flat $Ecm$ distribution, here denoted
by the Energy Density Function ($EDF$). From it follows that the 
Energy Density Function, the property of two colliding protons, is clearly not
equivalent to the $PDF$ property of a single non-reacting 
proton it does however 
serve as an essential input for the $EDF$ computation 
as shown further on.
The knowledge of the Energy Density as a function of the $Ecm$ 
is crucial for the transformation of the cross sections dependence on $Ecm$ 
from those calculated theoretically for the
free partons reaction  
to the ones where the partons are embedded in the colliding protons. 
Moreover, it is also
required for polarization measurements whenever they are averaged over
non-negligible $Ecm$ range to allow their comparison with
theoretical expectations. 

%\subsection{The need for the Energy Density Functions}
%In general theoretical calculations of cross sections
%and other properties of parton anti-parton  
%reactions
%leading to exclusive final states, such as
%$\bar q q \to W^+ W^-$, are carried out as a function of the
%parton anti-parton center of mass energy, $Ecm$. In the study of
%the same reaction in proton-proton collisions at a given
%center of mass energy $\sqrt{S}$, the 
%quantity $Ecm$ can be determined for each event from its final state
%momenta. The 
%question however is how the theoretical calculated cross section, e.g. 
%$\sigma(\bar q q\to W^+ W^-)$, is transformed when measured in
%$pp$ collisions. 
%For a specific parton anti-parton $Ecm$, there are 
%an infinite set of $x_k\times \bar x_k$ pairs that contribute 
%(where $x_k\ =\ E_k/E_{proton}$ is the fraction of the proton energy
%carried by the parton $k$, 
%when $x_k$ is a result of the first proton $PDF$ and $\bar x_k$ 
%is a result of the second proton $PDF$), 
%the sum of these contribution for
%each $Ecm$ value generates the Energy Density Function ($EDF$). 
%One should note that clearly the $EDF$
%information
%is needed for the cross section transformation from the 
%theoretical calculated reactions of free partons
%to the ones imbedded in the protons. It is however also
%required for a polarization measurement when it is averaged over
%a non-negligible $Ecm$ region so that it can be compared with
%the theoretical expectations. 
%Moreover, the calculation of the $EDF$ from the
%different $PDF$'s is not a straightforward process 
%as is shown further on.

\subsection{Evaluation of the Energy Density Functions}
One way to achieve an evaluation of the $EDF$ is 
via the use of a dedicated Monte Carlo (MC) program of $pp$ collisions
which generate individual parton anti-parton reactions  
and determine for each of them their $Ecm$ value. If a large enough event
sample is generated,
the relative repeated occurrence of each $Ecm$ value is proportional to
its corresponding $EDF$ estimate.
Such a dedicated MC program, which is time consuming and frequently
requires special development efforts, is often not readily available
for the particular reaction under study so that an analytical
evaluation of the $EDF$ is inevitable and hence is here further estimated.\\ 

If we denote by $S$ the center of mass energy squared 
of the colliding $pp$ system,
here set to the nominal LHC energy squared of (14 TeV)$^2$, 
and by $\hat s ~(=Ecm^2)$ the
center of mass energy squared of the initial interacting 
parton anti-parton pair, then
the following relation holds:
\begin{equation}
\hat s\ =\ x_kx_{\bar k} S\ ,
\end{equation}
where $x_k\ =\ E_k/E_{proton}$ is the fraction of the proton energy
carried by the parton $k$. For the different Parton Distribution
Functions, here denoted by $h_k(x_k)$, we use the  
CTEQ6.5M ones given in reference \cite{cteq}. 
For a fixed $S$, the probability 
$P(\hat s,S)d\hat s$ is given by
\begin{equation}
P(\hat s,S)d\hat s\ =\ \frac{\int^1_{min}dx_k \int^1_{min}dx_{\bar k}~
[h_k(x_k)\bar h_{\bar k}(x_{\bar k})\delta(x_kx_{\bar k}-\hat s/S)]d\hat s}
{\int_{\hat s_{min}}^{\hat s_{max}}d\hat s\int^1_{min}dx_k \int^1_{min}dx_{\bar k}~
[h_k(x_k)\bar h_{\bar k}(x_{\bar k})\delta(x_kx_{\bar k}-\hat s/S)]}\ ,
\label{eq_p}
\end{equation}
\begin{figure}[h]
\centering{\psfig{figure=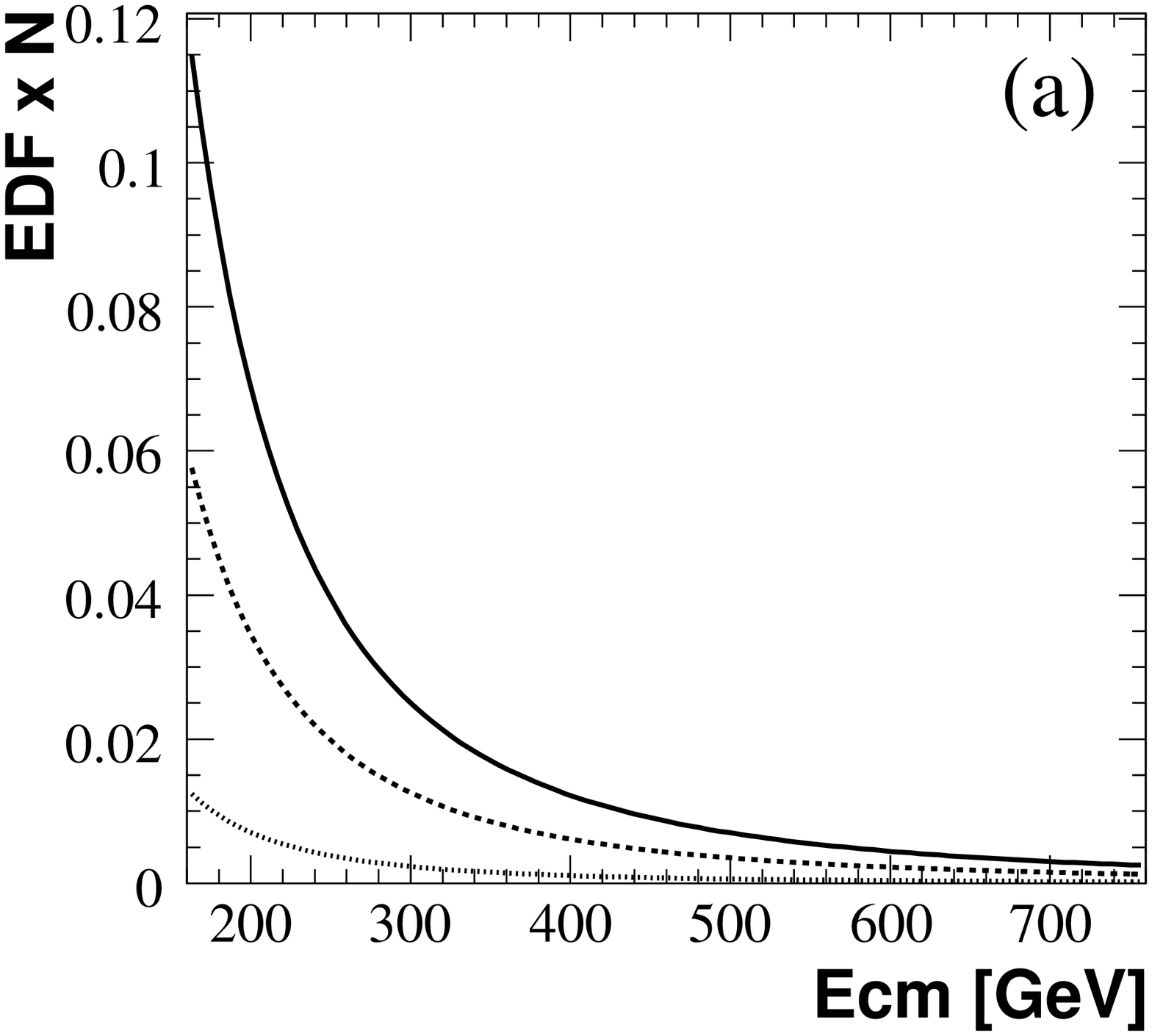,height=5.3cm}\   
\psfig{figure=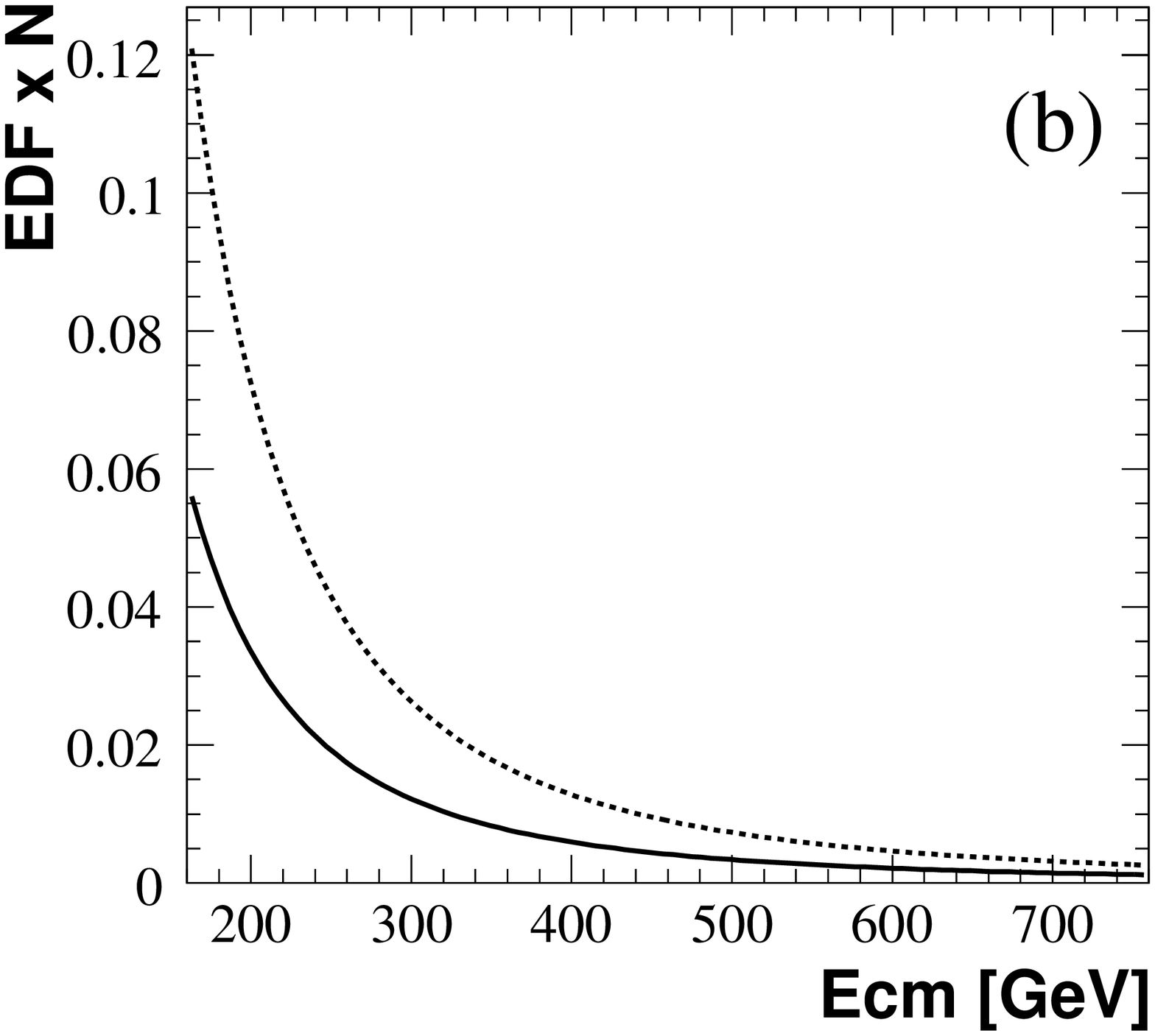,height=5.3cm} \ 
\psfig{figure=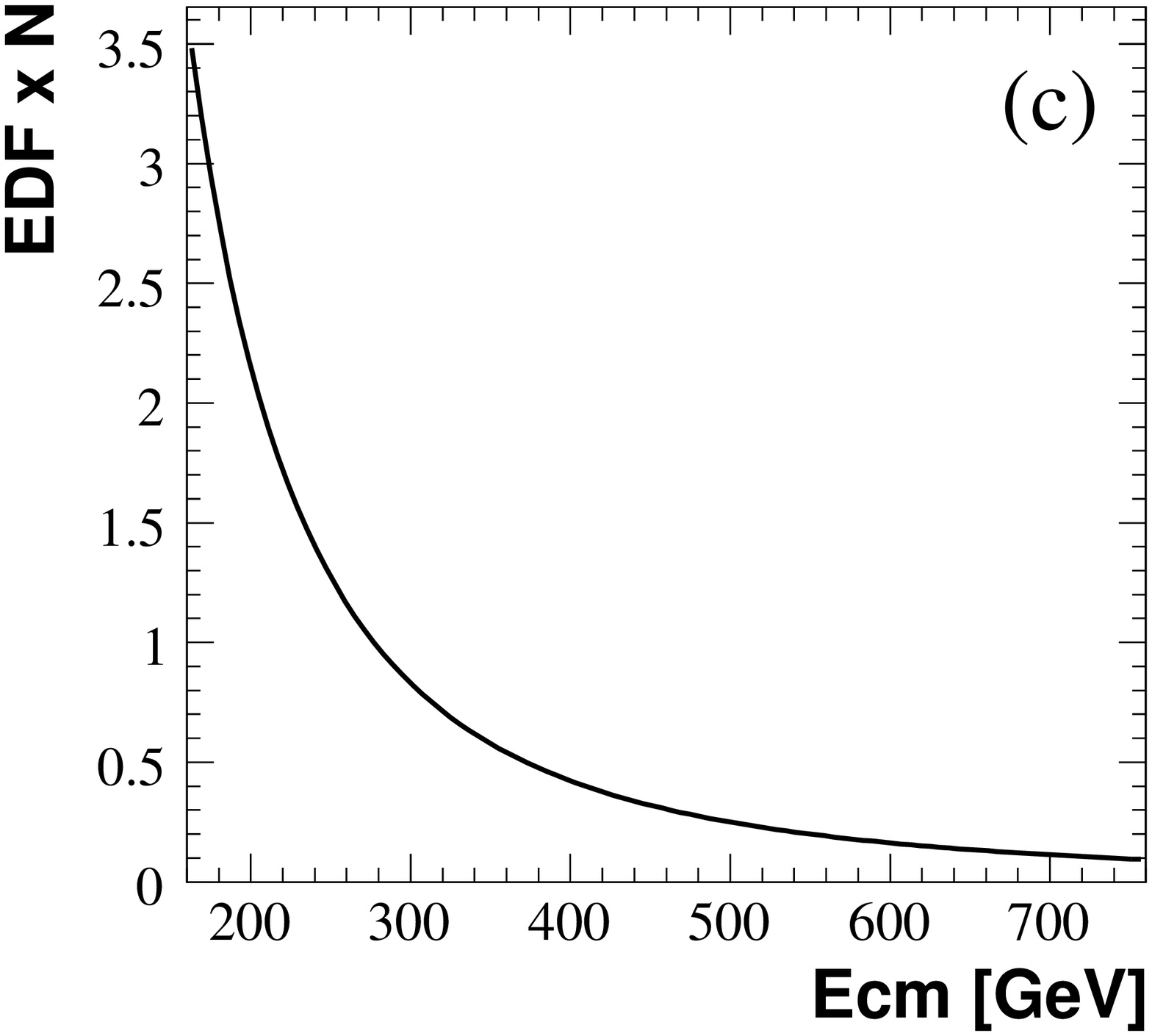,height=5.3cm}}
\caption{\small The unnormalized parton anti-parton Energy Density
Function, $EDF\times N$ as a function of $Ecm$
for $pp$ interactions at a $\sqrt{S}=14$ TeV 
given by Eqs. \ref{eqqq} and \ref{eqdu}.
(a):
The continuous, dashed and dotted lines are respectively 
the $EDF_{k \bar k}\times N_{k \bar k}$ dependence on $Ecm$ of the
$u\bar u$, $d\bar d$ and $s\bar s$ systems; 
(b): The continuous and dashed
lines are respectively the $EDF_{k \bar k}\times N_{k \bar k}$ dependence 
on $Ecm$ of the $d\bar u$ and $u\bar d$ systems; (c): The
$gluon-gluon$ $EDF_{g\bar g}\times N_{g\bar g}$.   
}  
\label{edf}

\end{figure}

\noindent
where the lower positive integration limits of $dx_k$ are set to very small, 
but non zero, values in order to avoid 
in the numerical calculations poles at $x=0$. To note is that these
two last formulae are also valid for the $gluon-gluon$ $(g\bar g)$
collisions. From the probability
distribution $P(\hat s,S)$ we derive the parton anti-parton center of
mass Energy 
Density Functions, $EDF(k \bar k)$, for $pp$ collisions at 14 TeV 
which can be parameterize above the $VV'$ threshold as 
\begin{equation}
EDF_{u \bar u} \simeq \frac{1}{N_{u\bar u}}\frac{3.9\times 10^4}{Ecm^{2.5}}; \ \ \  
EDF_{d \bar d} \simeq \frac{1}{N_{d\bar d}}\frac{1.956\times 10^4}{Ecm^{2.5}}; \ \ \ 
EDF_{s \bar s} \simeq \frac{1}{N_{s\bar s}}\frac{1.5\times 10^4}{Ecm^{2.75}},
\label{eqqq}
\end{equation}
where $1/N_{ij}$ are the normalization factors for the colliding
$k_ik_j$ partons
which depend on $\sqrt{\hat s_{min}}$ and $\sqrt{\hat s_{max}}$. 
The $EDF$ expressions for the quark anti-quark systems $d \bar u$
and $u \bar d$, as well as the $g \bar g$, are respectively given by
\begin{equation}
EDF_{d \bar u} \simeq \frac{1}{N_{d\bar u}}\frac{1.9\times 10^4}{Ecm^{2}}; \ \ \  
EDF_{u \bar d} \simeq \frac{1}{N_{u\bar d}}\frac{4.1\times 10^4}{Ecm^{2}};\ \ \
EDF_{g \bar g} \simeq \frac{1}{N_{g\bar g}}\frac{55\times 10^4}{Ecm^{2.35}}.   
\label{eqdu}
\end{equation}
The unnormalized $EDF$'s are shown in Fig. \ref{edf} as a function of $Ecm$.
These parameterized $EDF$ expressions are applied in the following sections to 
the di-boson production cross sections and in particular to the   
$q \bar q \to Z^0Z^0$ reaction discussed in Sec. \ref{sec_zz}.

\section{The $SM$ cross sections and polarizations}
\label{amplitudes}
For the calculations of the gauge boson pair production in quark 
anti$-$quark and $e^+e^-$ reactions we have utilized the $SM$
helicity amplitudes tables
given in Ref. \cite{nuss} 
for the $W^+ W^-$,  $W^{\pm} Z^0$ and
$Z^0 Z^0$ final states. 
We further specify by $\tau$ the 
%$\tau$ 
helicity states, $-1$, 0 and $+1$, of the 
outgoing bosons and by  
$\lambda$ we denote
the initial fermion helicity values of $\pm 1/2$. Furthermore, 
the angle $\theta_W$ represents   
as usual the electro-weak mixing angle, 
so that the vector ($a_Z$) and the axial vector ($b_Z$) couplings
of the $SM$ fermions ($f$)
to the $Z^0$ gauge boson are given by 
\begin{equation}  
a_Z\ =\ \frac{1}{4 sin\theta_W cos\theta_W}(r_3-4Qsin^2\theta_W)\ ;\ \
{\rm{and}}\ \ \ b_Z\ =\ \frac{1}{4 sin\theta_W cos\theta_W}r_3\ ,
\end{equation}
where $Q$  is the electric charge 
namely, $Q\ = \ 2/3, -1/3$ and $-1$ respectively for
the up, down quarks and the electron.
The weak isospin
projection $r_3$ of the fermions is equal to,  
$r_3\ =\ +1$ for the up quark while $r_3\ =\ -1$ is for the down quark and the
electron.  
After integrating over the 
azimuthal angle, the differential cross section of the process
$f \bar f \to VV'$ is
given in terms of the helicity amplitudes $F_{\lambda\lambda
'\tau\tau '}$ by
\begin{equation}
\frac{d\sigma}{d cos\theta}\ =\ \frac{C|\vec p|}{16\pi 
\hat s\sqrt{\hat s}}
\sum_{\lambda\lambda '\tau\tau '}|F_{\lambda\lambda '\tau\tau '}
(cos\theta)|^2\ ,
\label{xsec}
\end{equation}
where $\theta$ stands for
the production scattering angle defined in the
vector boson $VV'$ pair rest frame between the incident fermion 
and the final $V$ boson momenta. The average color factor $C$ is
equal to 1 for $e^+e^-$ and 1/3 for $q \bar q$ initial states.
For a given polar angle $cos\theta$
the single and joint longitudinal polarizations
$\rho_{00}(cos\theta)$ and $\rho_{0000}(cos\theta)$ are given by 
\begin{equation}
\rho_{00}(cos\theta)\ =\ 
\frac{\sum_{\lambda\lambda '\tau '}|F_{\lambda\lambda ' 0\tau '}(cos\theta)|^2}
{\sum_{\lambda\lambda '\tau\tau '}|F_{\lambda\lambda '\tau\tau '}
(cos\theta)|^2}
\ \ \ \ \ {\rm{and}}\ \ \ \ \  
\rho_{0000}(cos\theta)\ =\ 
\frac{\sum_{\lambda\lambda '}|F_{\lambda\lambda ' 0 0}(cos\theta)|^2}
{\sum_{\lambda\lambda '\tau\tau '}|F_{\lambda\lambda '\tau\tau '}
(cos\theta)|^2}\ .
\label{rho}
\end{equation}
Similar expressions define the other $\rho$ elements like the
$\rho_{++--}$ and $\rho_{--++}$.\\ 

In the following sections we will address our calculations to the
longitudinal polarizations averaged over $cos\theta$  
which we will
refer to simply as $\rho_{00}$ and $\rho_{0000}$. These polarizations
can be estimated from the measurement of the $cos\theta_f$ 
distributions where
$\theta_f$ is the angle between the decay fermion direction in its 
parent gauge boson rest frame and the gauge boson direction in the 
di-boson  
center of mass system. Two leading methods, which are
described e.g. in Ref. \cite{opal_w_triple},
are used to extract the polarization from the $cos\theta_f$
distributions. 
The first is by fitting it to an expression for 
$d\sigma/dcos\theta_f$ given in terms of the helicity states and the
second one by applying the
$\Lambda_{ij}$ helicity projection operators.
 
\section{The reactions $f \bar f \to V V'$ within the $SM$}
\label{sec_ww}
\subsection{ $\bar f f \to  W^+ W^-$} 
The reaction $e^+ e^- \to  W^+ W^-$ has been studied in LEP2 in the
center of mass energies up to $\sim$210 GeV \cite{mele} using unpolarized 
electron-positron
beams.
The total and differential cross sections 
have been measured and found to be consistent with the $SM$ expectations.
%In addition $\rho_{00}$, the fraction of the longitudinal polarization 
%of the single 
%$W$ boson \cite{delphi_w_pol,l3_w_pol,opal_w_pol}  
%and the fraction of the joint $W^+$ and 
%$W^-$ longitudinal polarizations, $\rho_{0000}$,  
%were also evaluated. 
The $\rho_{00}$ measurements results
are summarized in Table \ref{tab_ww_pol} where they are found to be
consistent with the $SM$ expectations as 
estimated from our helicity amplitude calculations. The OPAL 
collaboration reported
also on 
a $\rho_{0000}$ measurement \cite{opal_w_triple} in the reaction
$e^+e^- \to W^+W^- \to \ell \nu q \bar q$ at the center of mass energy of ~189 GeV. 
Their joint polarization value of 
$0.201\pm 0.072(stat)\pm 0.017(syst)$ is clearly higher, 
but still essentially consistent within its large errors, 
with our calculated $SM$ expectation of 0.095 and that quoted by OPAL 
of 0.086$\pm$0.008.  
\renewcommand{\arraystretch}{1.1}
\begin{table}[h]
\caption{\small The LEP2 measurements of the $W$ longitudinal
polarization in $e^+ e^- \to W^+ W^-$ reaction at 
an average center of mass energy of 196.6 GeV.
}
 \begin{center}
    \begin{tabular}{||l|c|c||}
      \hline\hline
Experiment& $\rho_{00}$ & Reference\cr
\hline\hline
DELPHI & 0.249$\pm$0.045$\pm$0.022& \cite{delphi_w_pol} \cr
L3 & 0.218$\pm$0.027$\pm$0.016 & \cite{l3_w_pol}\cr
OPAL &0.239$\pm$0.021& \cite{opal_w_pol}\cr
\hline
$SM$ expectation & 0.22 & Present work\cr
\hline\hline
\end{tabular}
\end{center}
\label{tab_ww_pol}
\end{table}
\renewcommand{\arraystretch}{1.0}

\begin{figure}
\centering{\psfig{figure=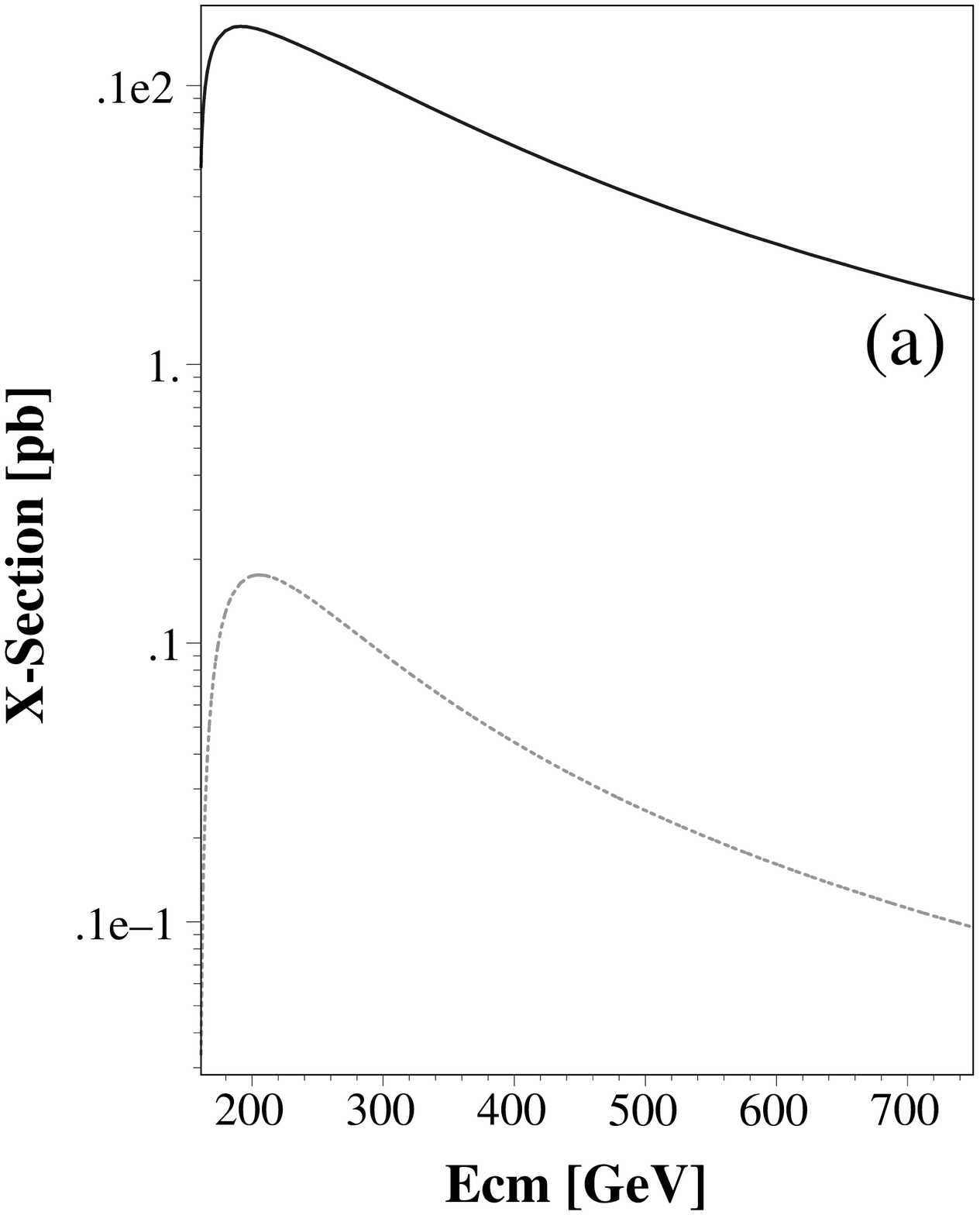,height=8cm} \ \
  \ \ \ \ 
{\psfig{figure=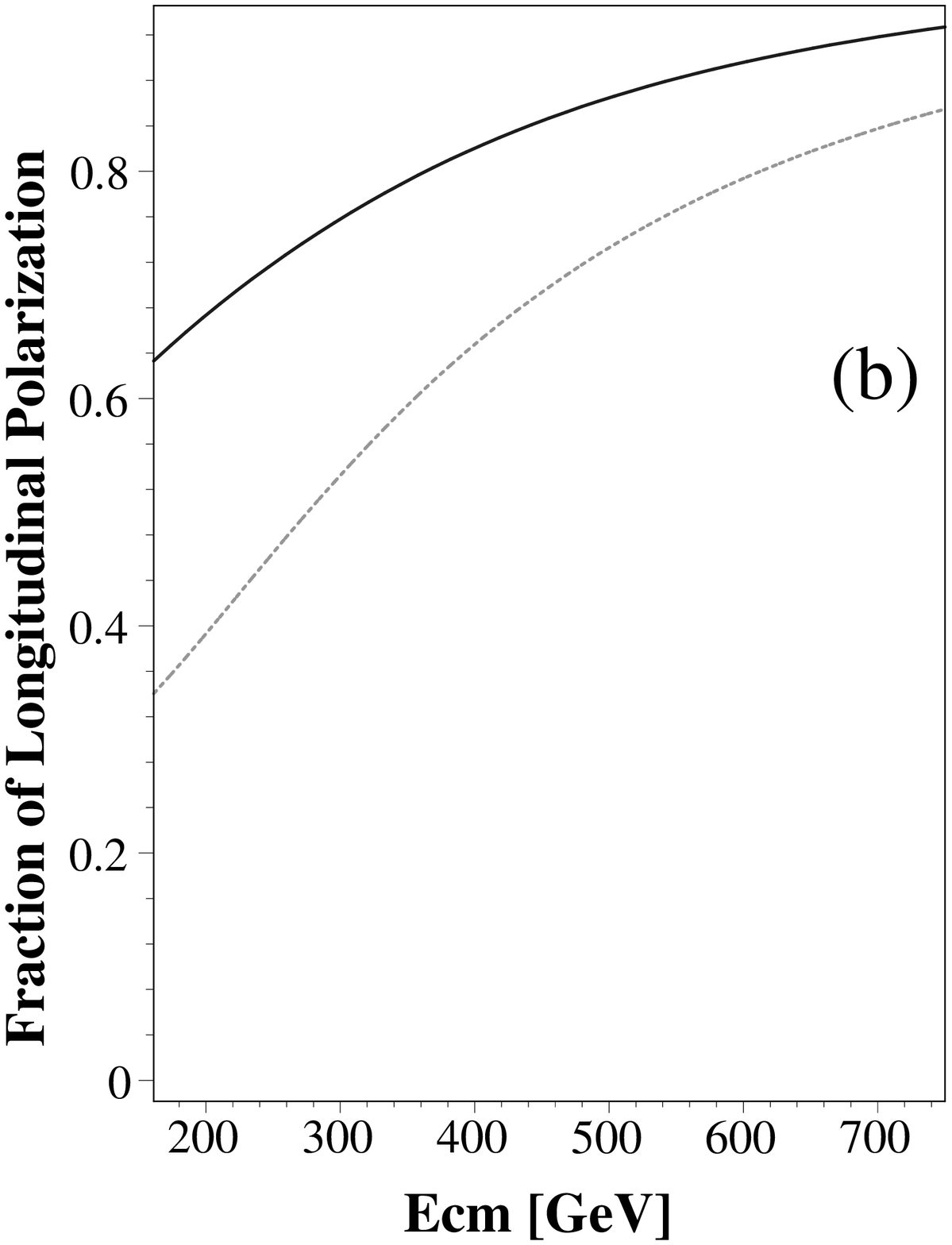,height=8cm}}} 
\caption{\small  
(a): $\sigma(e^+e^- \to W^+W^-)$ as a function of the $WW$ $Ecm$. 
The solid line
is for left-handed electrons and the dashed line is for 
right-handed electrons. 
(b): The $W$ longitudinal polarization as a function of the $WW$ $Ecm$ for
a right-handed electron initial state. The solid and dashed line
represent respectively 
the single and joint longitudinal polarization of the
$W^+$ and $W^-$ bosons.
} 
\label{ee-ww-pol}
\end{figure}

In the future ILC the electron and positron beams are
planned to be longitudinal polarized. As a consequence,
$\sigma(e^+ e^- \to W^+ W^-)$ and the expected longitudinal  
polarizations of the final state $W^+$ and $W^-$ will depend on
the polarization configuration of the initial state fermions.
For an unpolarized positron and a right-handed and left-handed
electron, the  $e^+ e^- \to W^+ W^-$ cross sections as
a function of the $W^+W^-$ $Ecm$ are shown in Fig. \ref{ee-ww-pol}a.  
\begin{figure}
\centering{\psfig{figure=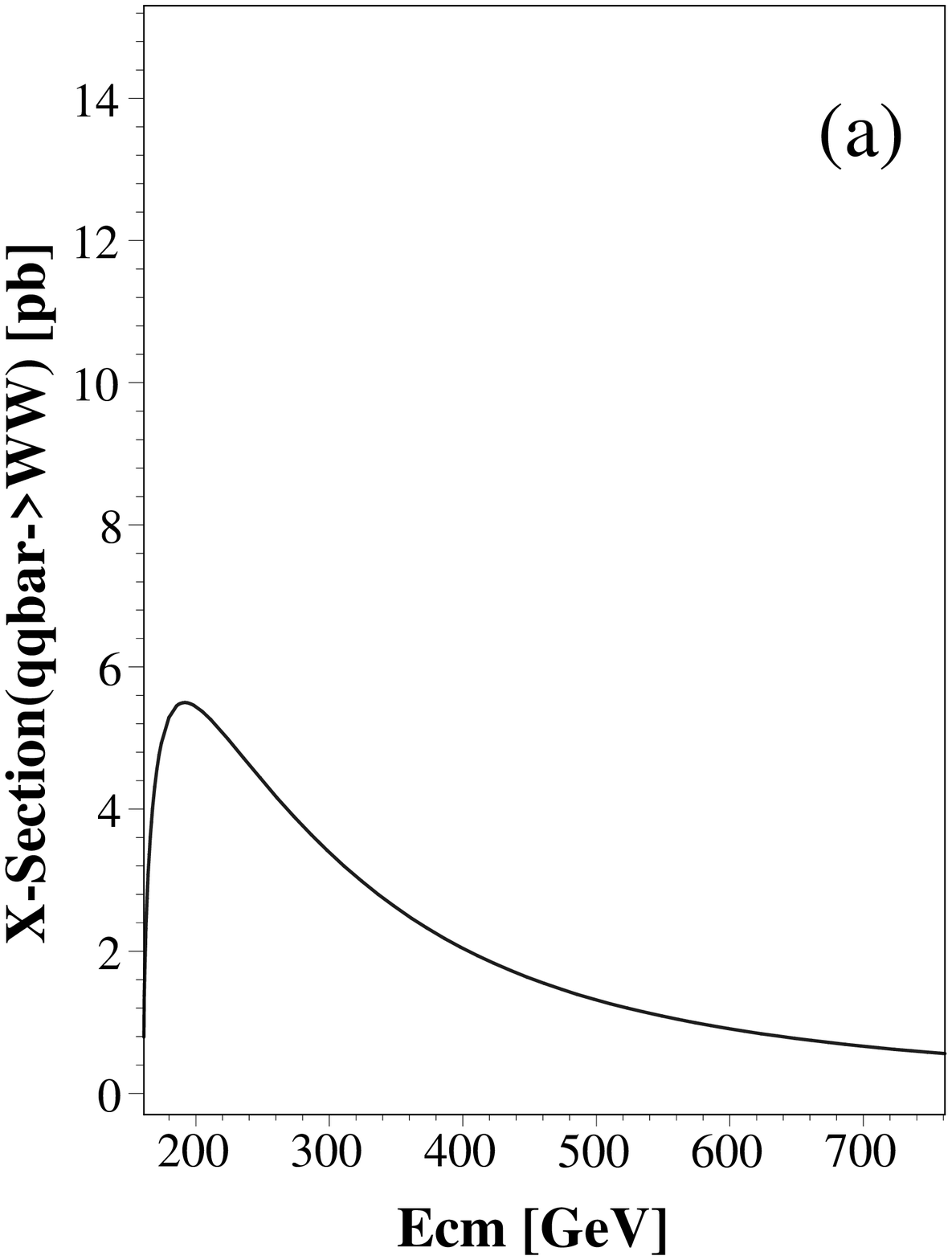,height=7cm} \ \   
{\psfig{figure=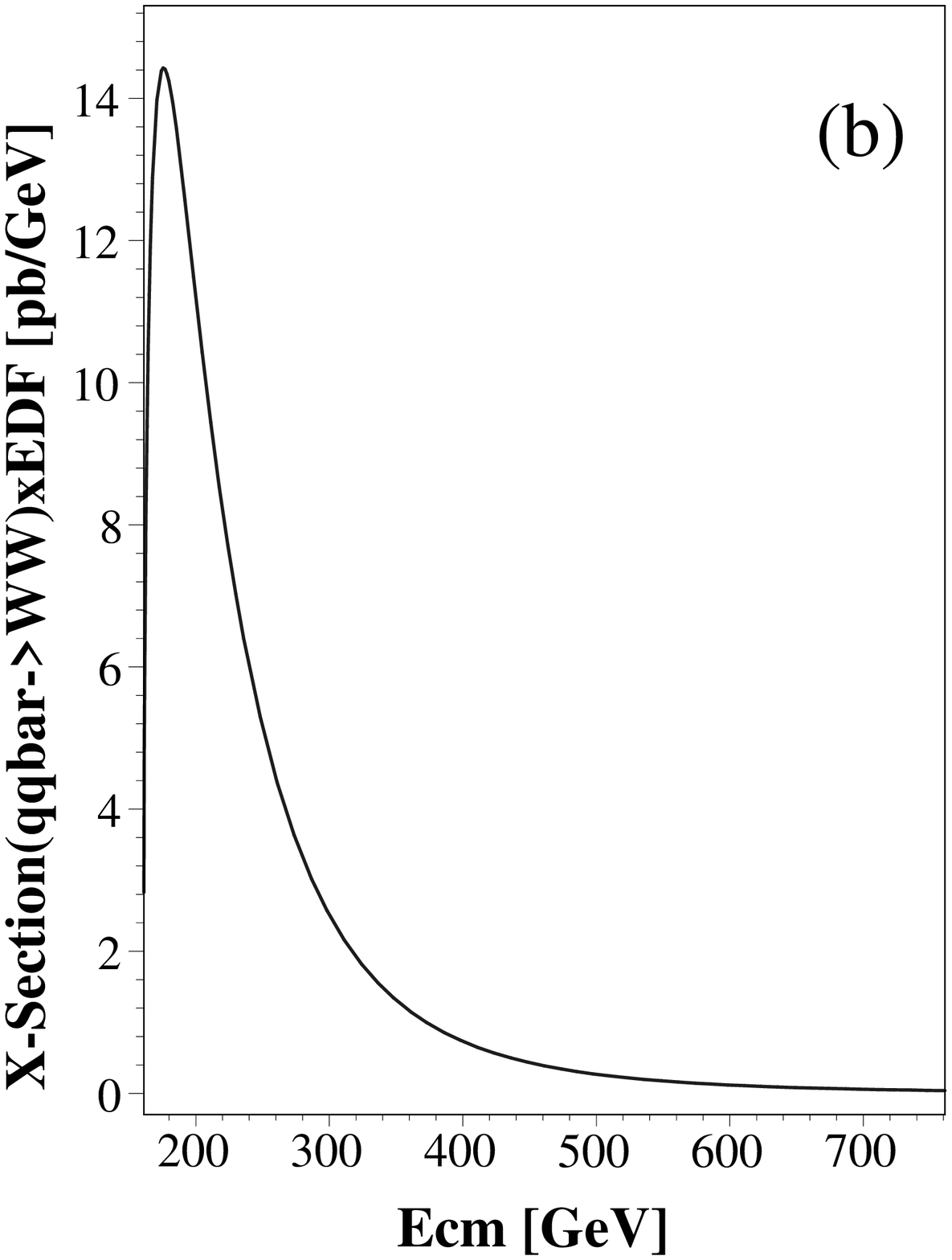,height=7cm}}\ \ 
{\psfig{figure=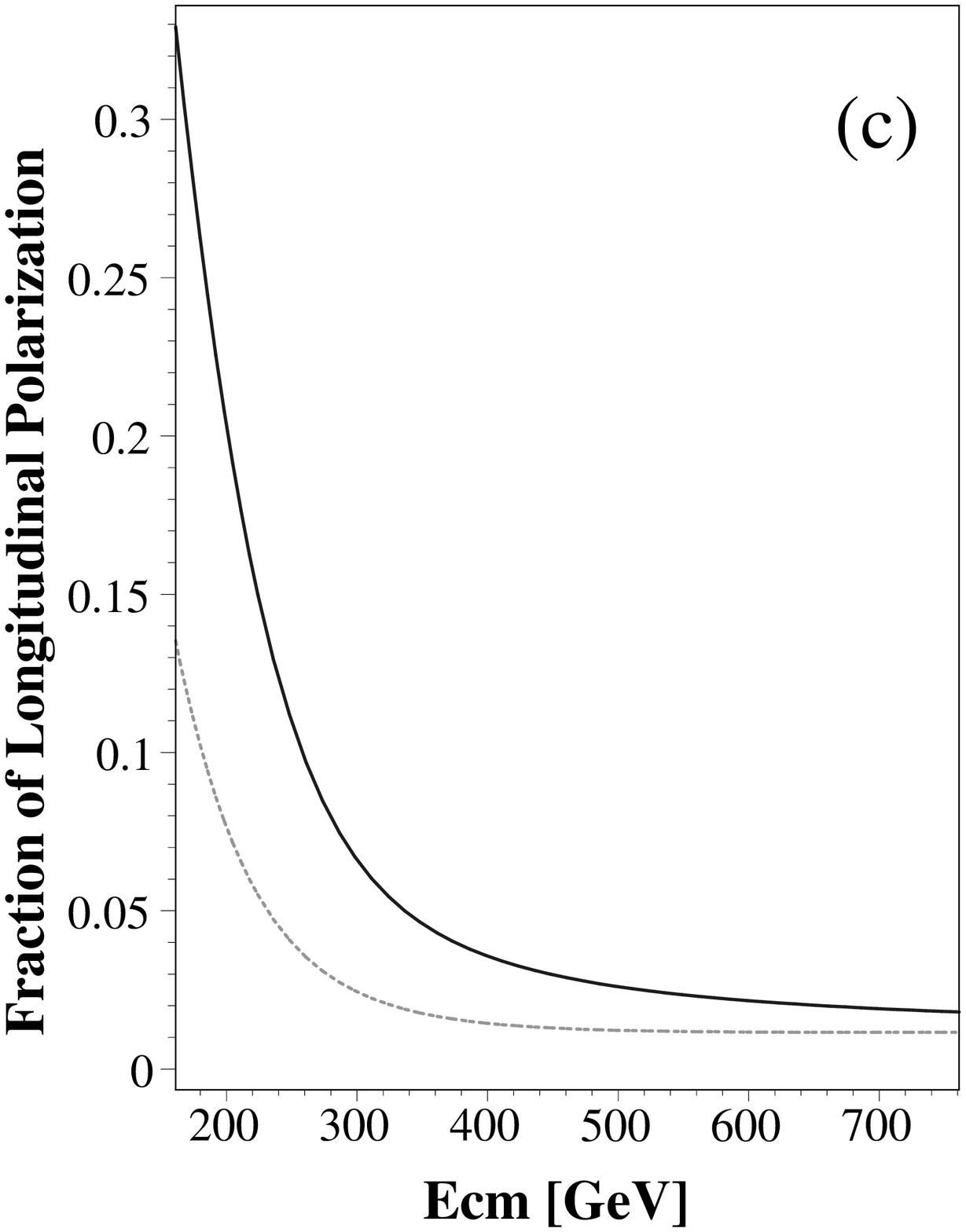,height=7cm}}}
\caption{\small (a): $\sigma(\bar q q \to W^+ W^-)$ as a function of 
the $WW$ $Ecm$;
(b): $\sigma(pp \to W^+ W^-)$ at $\sqrt{S_{pp}}=$14 TeV as a function of
the $WW$ $Ecm$ obtained via $EDF$ from $\sigma(\bar q q \to W^+ W^-)$  
and normalized to the area under the cross section shown in (a);
(c): Fraction of the longitudinal polarization as a function
of the $WW$ $Ecm$. The solid line is the single $W$ polarization 
$\rho_{00}$, and the
dashed line is the joint $W^+ W^-$ polarization, $\rho_{0000}$.    
} 
\label{dd-ww-xsec}
\end{figure}
To note is that the cross section of the right-handed electron
is much smaller than that of the left-handed electron which
approaches the value of the cross section for unpolarized beams.
Also of interest is the behaviour of the single and joint $W$
longitudinal polarizations (see  Fig. \ref{ee-ww-pol}b)
produced by the right-handed electron, previously
also dealt with in Ref. \cite{poulose},
which increases, rather than decreases, with energy and approach 
asymptotically the value of $\sim$$100\%$.\\

In the LHC, where the  proton beams are unpolarized, 
the production of the final state 
$W^+W^-$ is dominated by unpolarized $u \bar u$ and $d \bar d$ 
and to a lesser extent by the $s \bar s$ collisions 
with the relative ratios of 
$0.63,\ 0.316 \ {\rm{and}}\ 0.054$ respectively, 
obtained from the $PDF$ formulae.
The contribution
of the process $gluon\ gluon \to W^+W^-$ is estimated 
\cite{csc} to be of the order of $5\%$.
The $\bar q q \to W^+W^-$ cross section as a function of $Ecm$ 
is shown
in Fig. \ref{dd-ww-xsec}a where we sum over the $u\bar u$,
$d\bar d$ and $s\bar s$ initial states weighted according to
their occurrence in $pp$ collisions.  
In Fig. \ref{dd-ww-xsec}b is shown the same 
cross section as it will be observed in $pp$ collisions at 14 TeV
by applying the relevant $EDF$.
The $Ecm$ dependence of $\rho_{00}$ and $\rho_{0000}$ are 
illustrated in Fig. \ref{dd-ww-xsec}c where one observes that 
both decrease with energy and approach 
a nearly common flat value of
$\sim$0.02. In the event that the $W^+ W^-$ pair is produced in 
$e^+e^-$ collisions, like in LEP2, the $Ecm$ is known from the initial state, 
and the $W$ longitudinal polarization measurement does not present
any difficulty.
In the LHC however, even the single $W$ polarization measurement is not a
straightforward task
since the $Ecm$ value of each $W^+W^-$ event is not known from the initial
$\bar q q$ state and in most cases also not from the final state. 
As is well known, for the polarization measurement 
one has first to identify the $W^+W^-$ event 
which is dealt with in length in reference \cite{csc} for their decay
to the $\ell\nu\ell\nu$ and  
$\ell\nu\ q_1\bar q_2$ final states.
The following
requirement for the polarization measurement is the necessity 
to transform the $W^+W^-$ pair, including  their  decay products, to 
the $W^+W^-$ center of mass
system. 
This requirement should a priori
be more accessible 
in the process $W^+W^- \to q_1q_2\ q_3q_4$ than in their
leptonic decay modes. However, so far
an algorithm to identify in $pp$ collisions  
the $W^+W^-$ system via its fully hadronic decay configuration  
is still missing. Finally for the longitudinal
polarization measurement the 
polar angle of the $W$ decay products has to be determined
in the center of mass of the parent $W$ boson, a procedure 
already utilized previously \cite{opal_w_triple}. 
 
\subsection{The reaction $f' {\bar f} \to W^{\pm} Z^0$}
\label{sec_wz} 
Even though the $p p \to W^{\pm}
Z^0$ production cross section at 14 TeV is smaller by some $43\%$ than 
that of the $p p \to W^+ W^-$, it has the
advantage that the
$Z^0$ 2-body decay modes, $\mu^+ \mu^-$ and $e^+ e^-$, provide
a simple and 
clear identification and allow the measurement
of its longitudinal polarization inasmuch that the $W^{\pm}Z^0$ center of
mass system is accessible. Here we note in passing that
the final states $W^{\pm} Z^0$ cannot be reached
in $e^+ e^-$ collisions.\\ 
\begin{figure}
\centering{\psfig{figure=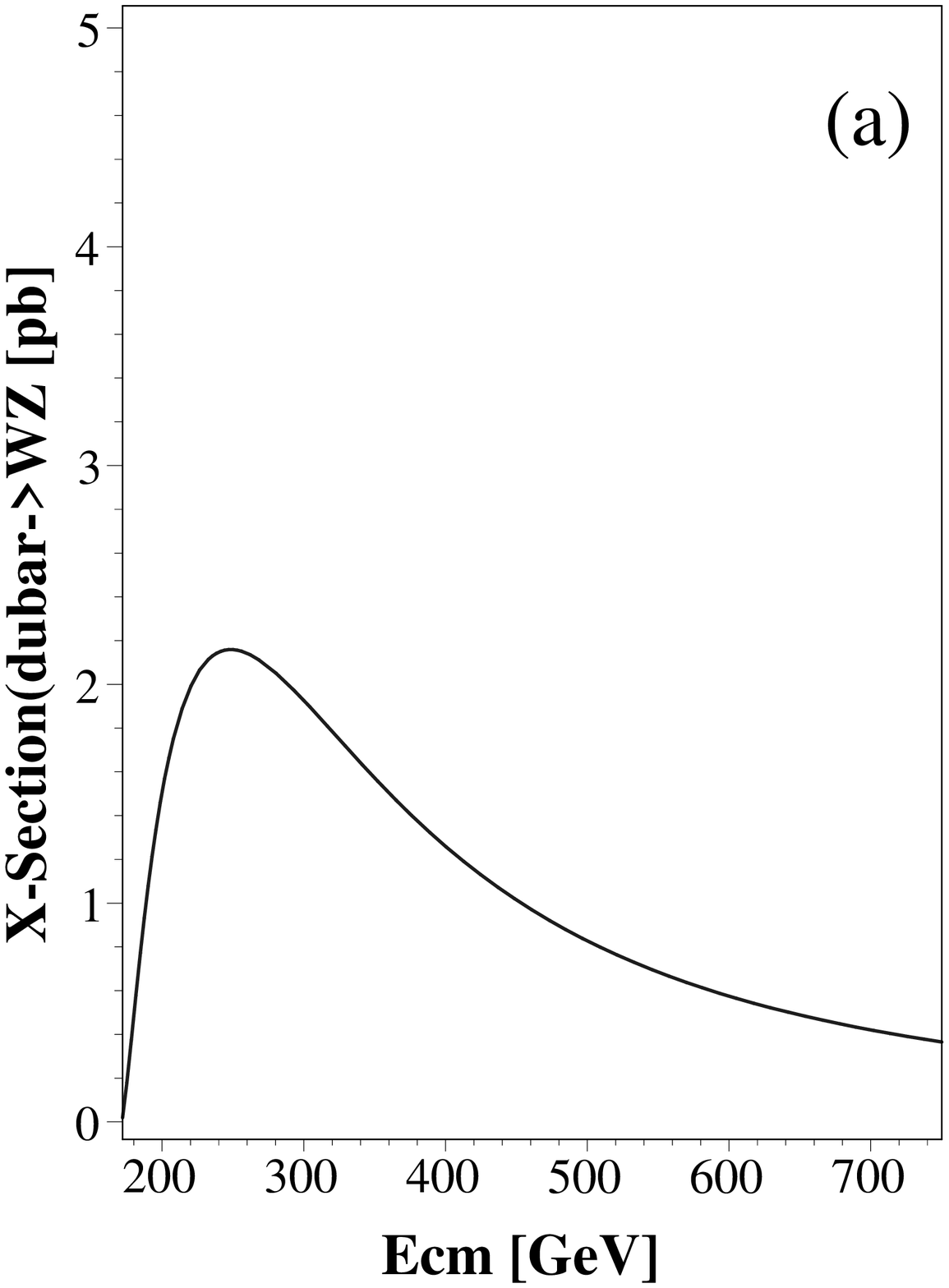,height=7.1cm} \ \  
{\psfig{figure=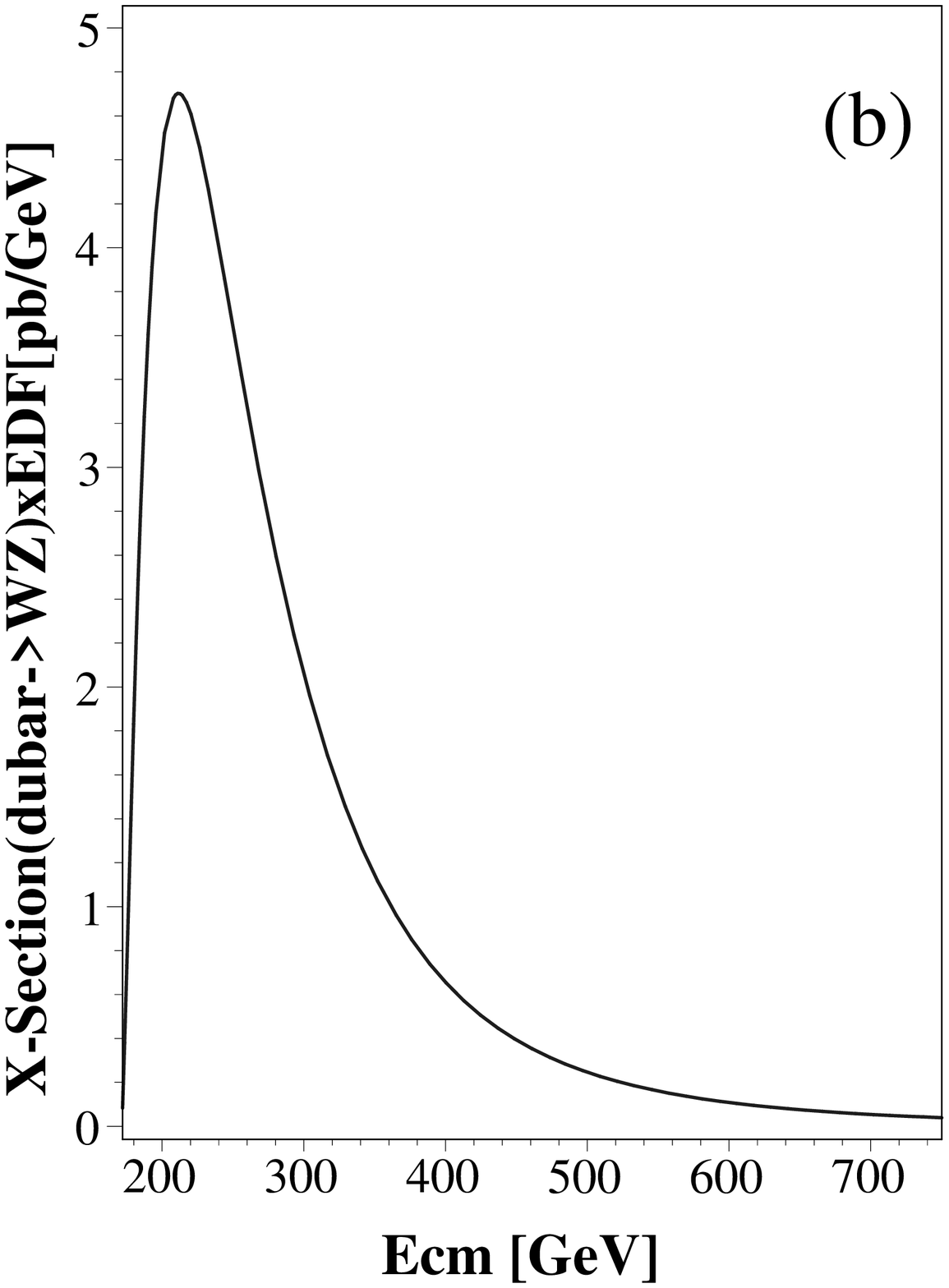,height=7.1cm}}\ \ 
{\psfig{figure=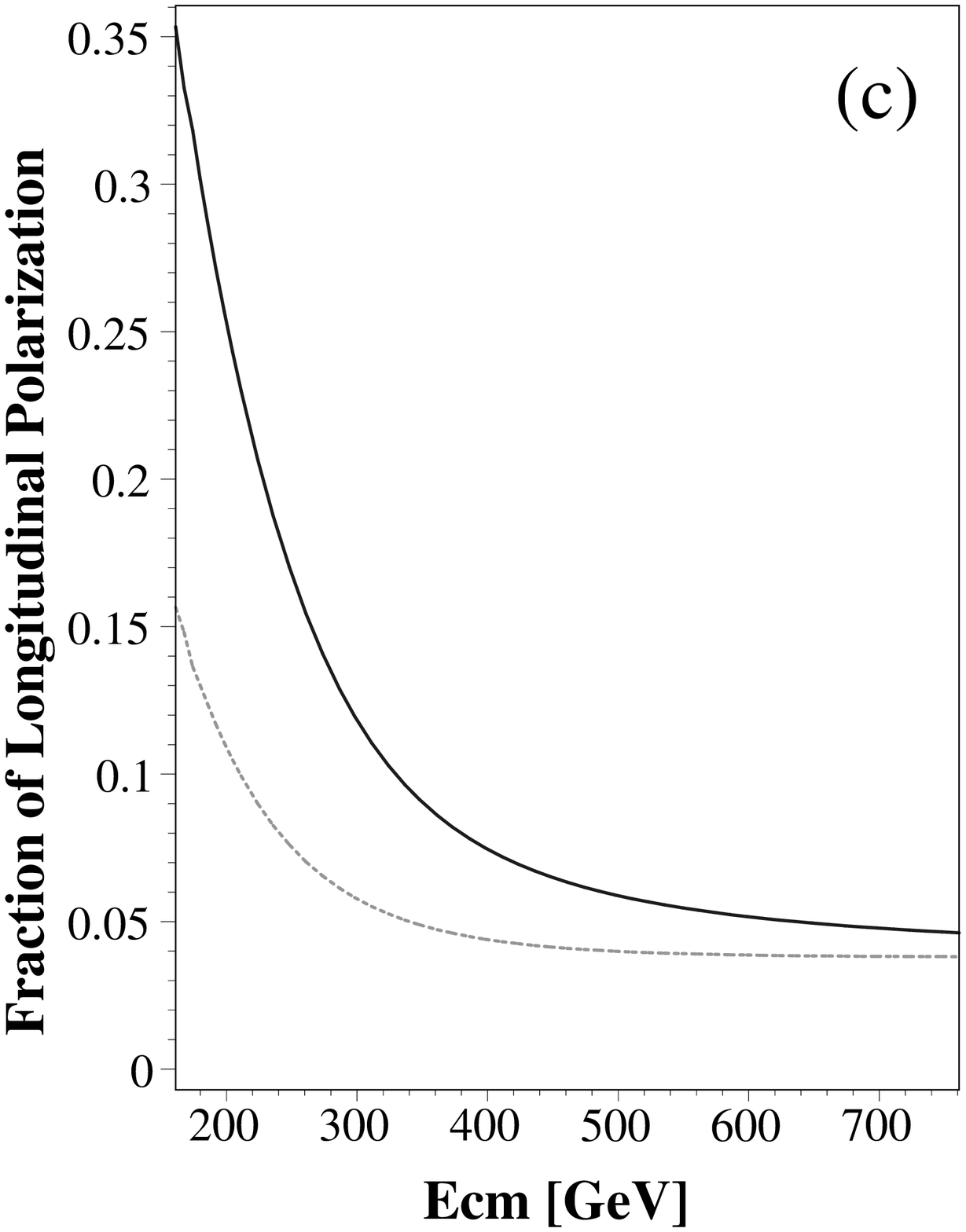,height=7cm}}}
\caption{\small (a): $\sigma(d\bar u \to W^-Z^0)$ as a function
of the $WZ$ $Ecm$; (b): $\sigma(pp \to W^-Z^0)$ at $\sqrt{S_{pp}}$=14 TeV 
as a function of the $WZ$ $Ecm$ evaluated from 
$\sigma(d\bar u \to W^-Z^0)$
via
the $EDF$ and normalized to the area underneath 
the cross
section given in (a);
(c): The longitudinal polarization 
as a function of the $WZ$ 
$Ecm$. The solid line is the single $Z^0$ polarization $\rho_{00}$
and the dashed line is the joint $W^-Z^0$ polarizations $\rho_{0000}$.    
} 
\label{dubar-wz-xsec}
\end{figure}
 
In Figs. \ref{dubar-wz-xsec}a and \ref{dubar-wz-xsec}b 
our calculated $SM$ expectations are shown for the 
$d \bar u \to W^-Z^0$ cross section as a function of $Ecm$
without and with the $EDF$ inclusion.
The single $\rho_{00}
(Z^0)$ and the joint $\rho_{0000}(W^-Z^0)$ longitudinal polarizations 
are presented in Fig. \ref{dubar-wz-xsec}c
as a function of $Ecm$. 
Like in the case of the reaction $\bar f f \to W^+W^-$,  
with the increase of energy   
$\rho_{00}$ and $\rho_{0000}$ are seen to approach each other 
to reach a low flat plateau. The experimental determination
of the cross section and longitudinal polarization as 
a function of energy requires the possibility to calculate the $Ecm$ 
of the $WZ$ system. This requirement is
satisfied if one is able to identify the $W^{\pm}$ boson through its decay
into $q \bar q'$ i.e., in a 2-jet configuration. On the other hand in the
situation where the $Z^0$  
decays to $e^+ e^-$ or $\mu^+ \mu^-$ and the $W^{\pm}$ decays to
$e^{\pm} \nu$ or $\mu^{\pm} \nu$ one finds in general two solutions
for $Ecm$ which hinders a polarization
estimation. 
     
\subsection{The reaction $f {\bar f} \to Z^0Z^0$}
\label{sec_zz}
\begin{figure}
\centering{\psfig{figure=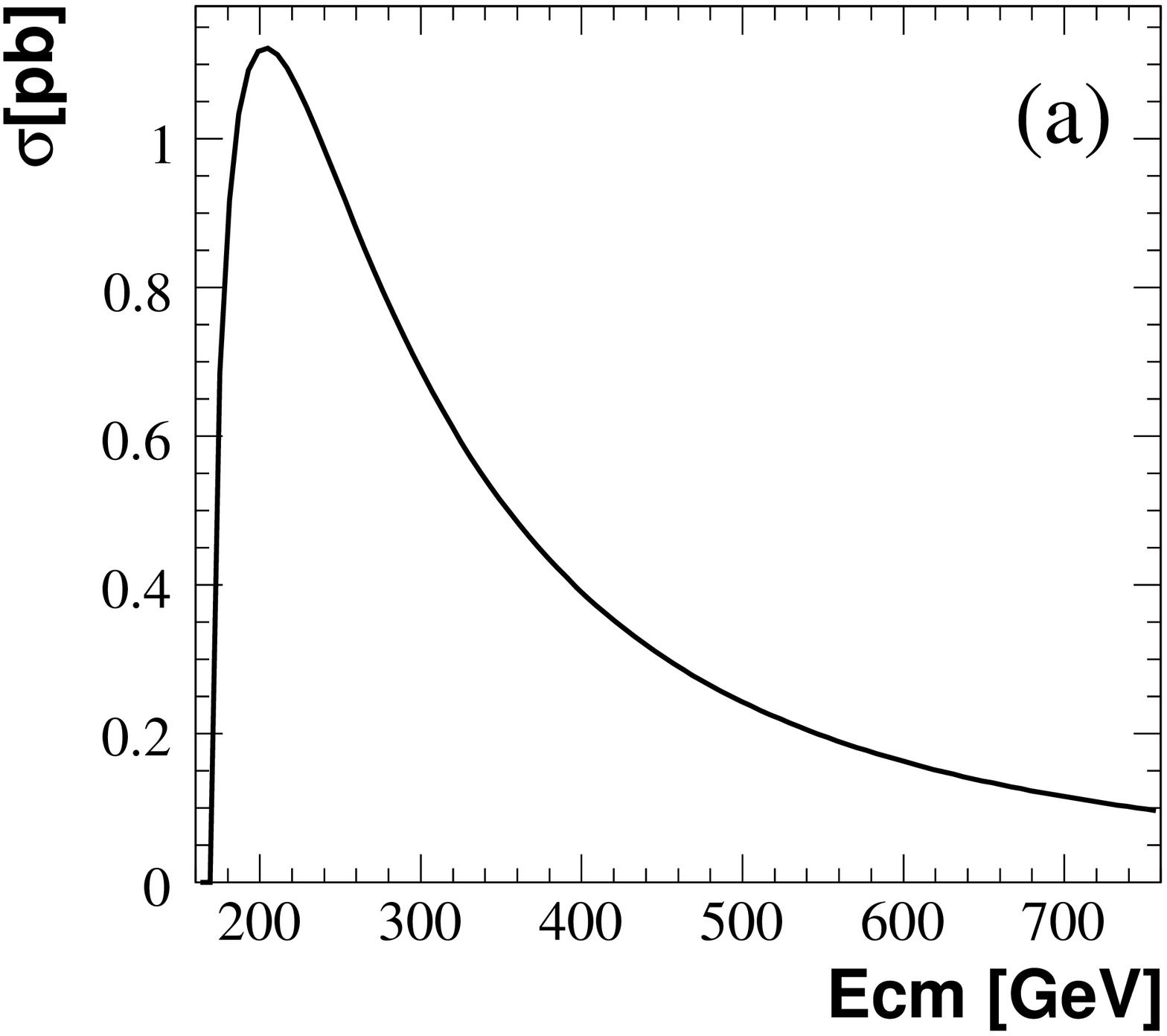,height=8cm} \ \   
\psfig{figure=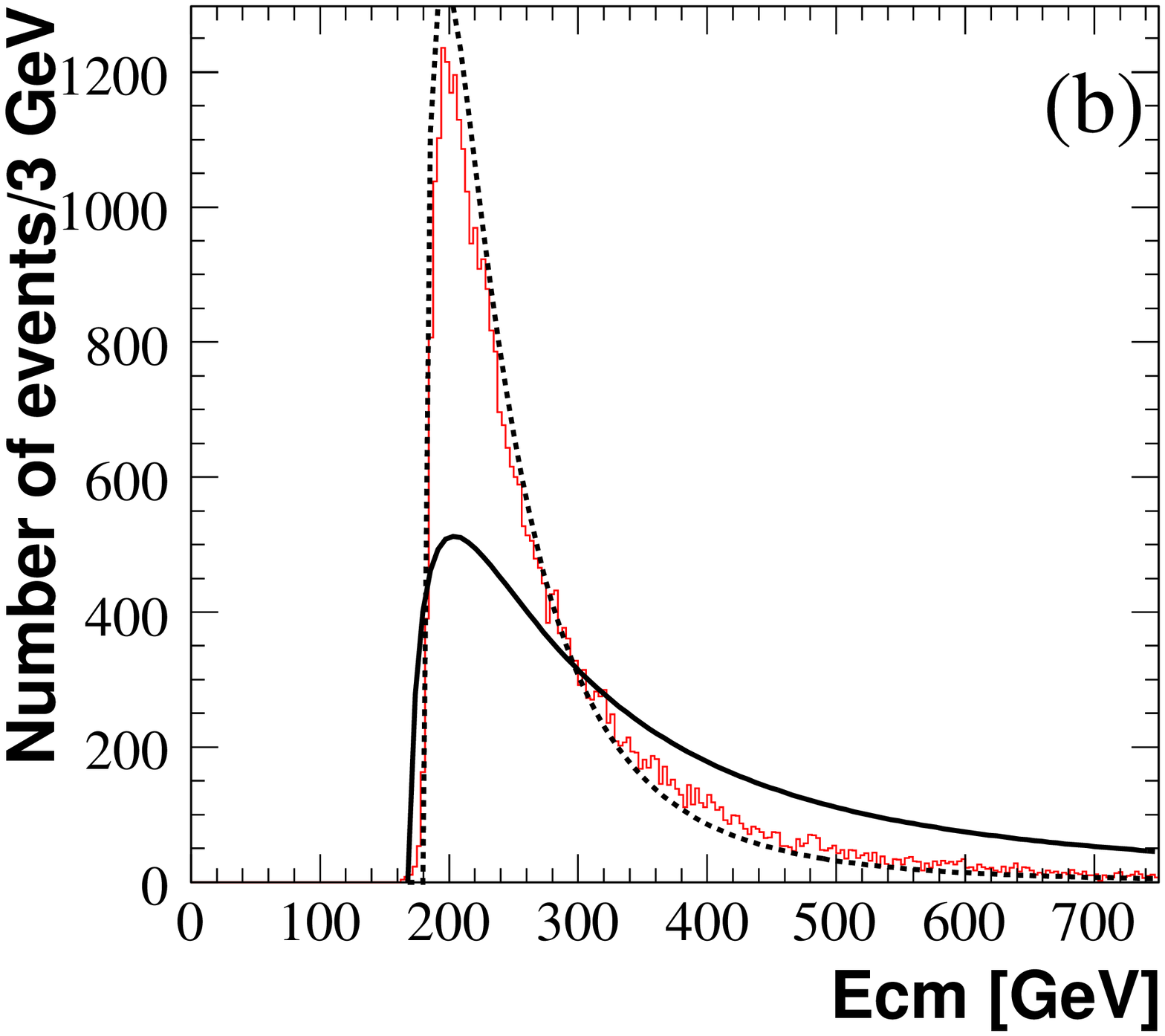,height=8cm} \\ 
\psfig{figure=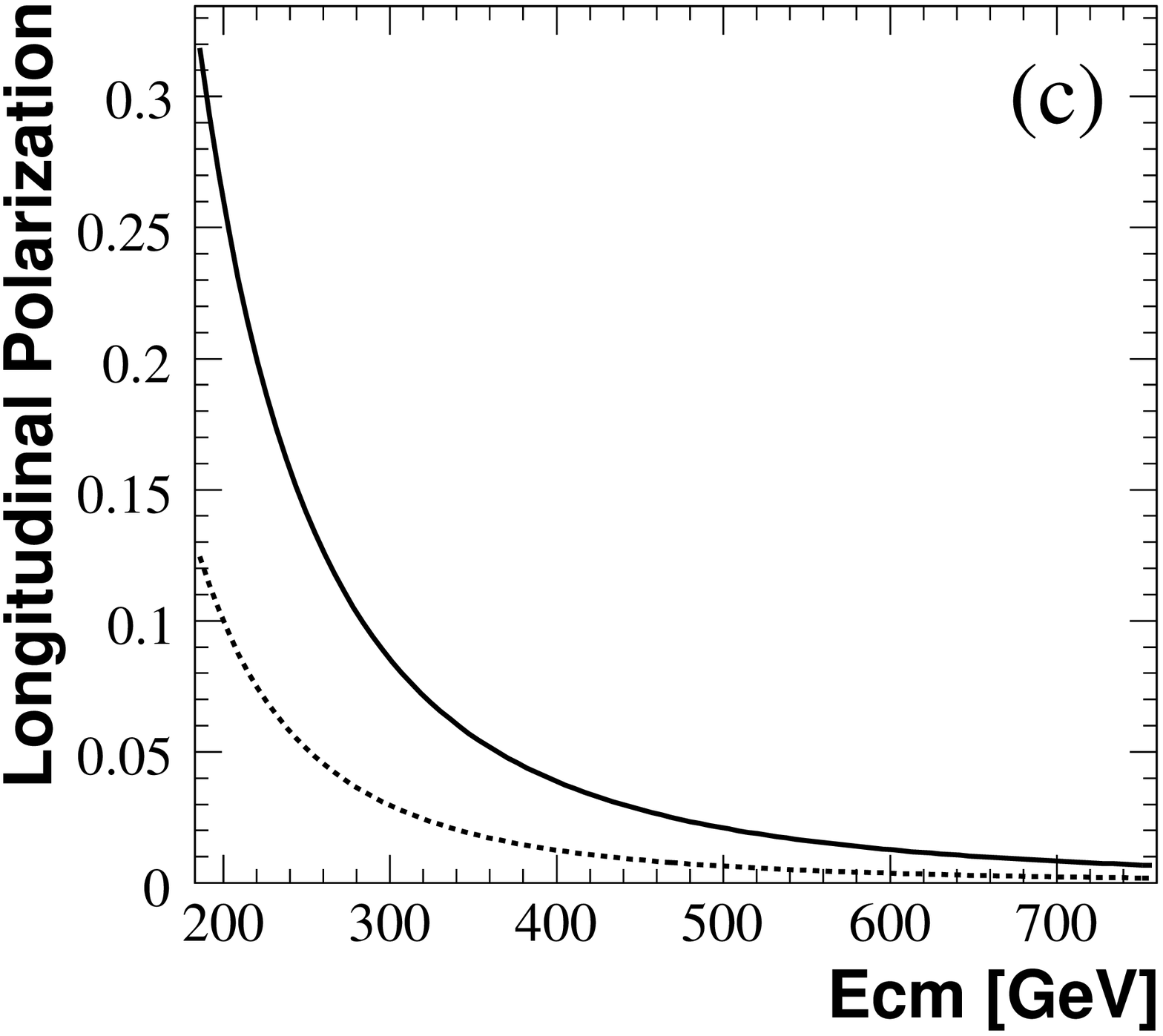,height=8cm} \ \
\psfig{figure=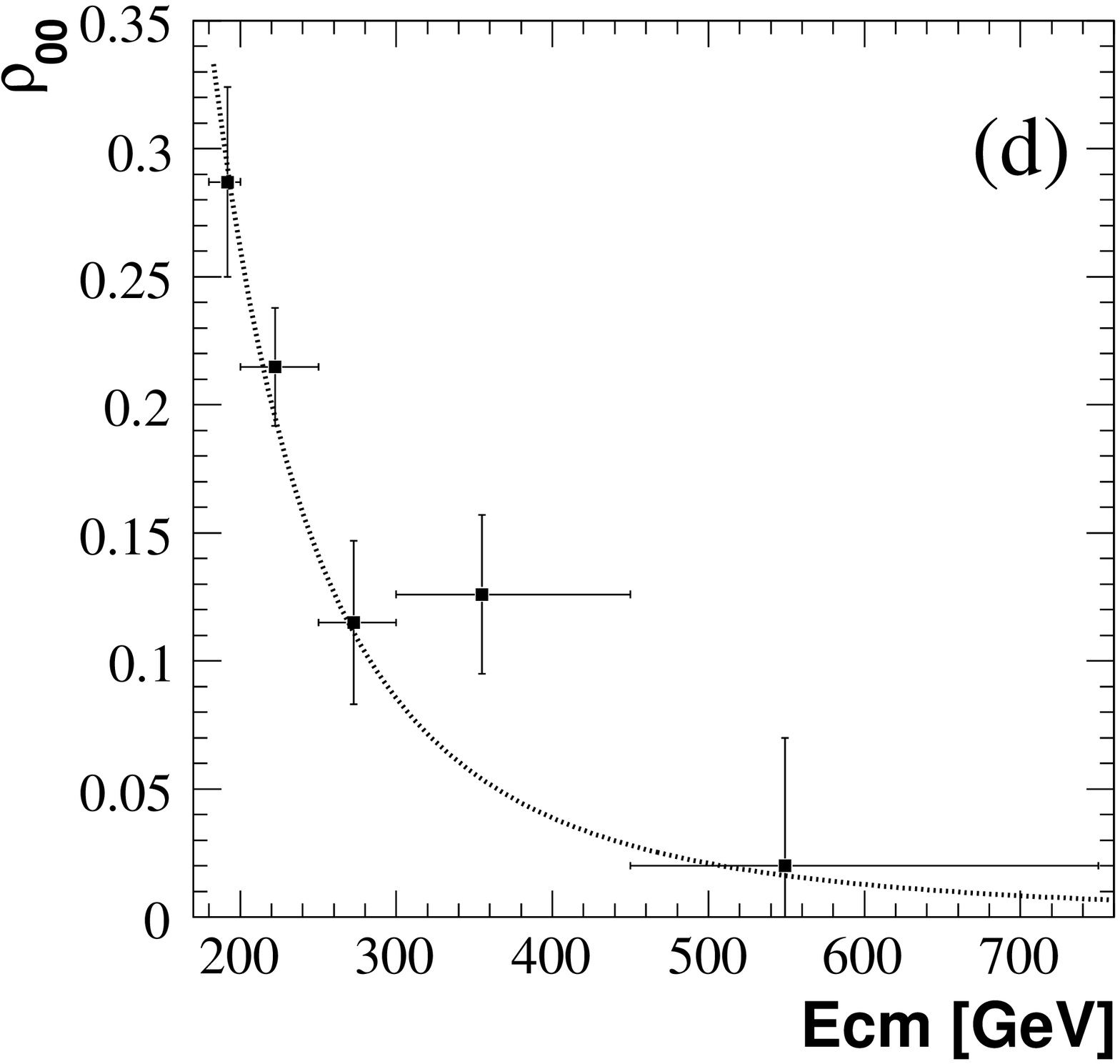,height=7.9cm}}
\caption{\small (a): $\sigma(q\bar q \to Z^0Z^0)$ 
as a function of the $ZZ$ $Ecm$; (b): Monte Carlo (MC) 
generated $pp \to Z^0Z^0$ 
events at $\sqrt{S_{pp}}$=14 TeV as a function of the $ZZ$ $Ecm$.
The solid and the dashed lines are respectively
$\sigma(q\bar q \to Z^0Z^0)$ and the corresponding 
$\sigma(pp \to Z^0Z^0)$ at $\sqrt{S_{pp}}$=14 TeV obtained by using the
$EDF$. Both lines are normalized so
that the area underneath them is equal to that
of the MC data histogram;   
(c): The calculated longitudinal polarization 
as a function of the $ZZ$ $Ecm$. The solid and dashed lines are respectively
the single, $\rho_{00}$, and the joint, $\rho_{0000}$, 
longitudinal polarizations;
(d): The measured $\rho_{00}$ from a MC generated $pp
\to Z^0Z^0$ data sample, equivalent to $pp$ integrated luminosity of
\mbox{$\simeq$200 $fb^{-1}$}, compared to the $SM$ expectation given
by the continuous line.     
} 
\label{dd-zz-xsec}
\end{figure}
The $Z^0Z^0$ system, in contrast to the $W^{\pm}Z^0$ and $W^+W^-$ 
final states, allows in most cases a straight forward determination of the 
center of mass energy event by event, in particular in its decay to
two pairs of charged electrons or muons. 
Our $SM$ helicity amplitude calculation 
of $\sigma(q \bar q \to
Z^0Z^0)$ as a function of $Ecm$, 
where we sum over the  $u\bar u$, $d\bar d$ and $s\bar s$ initial states, 
is shown in Fig. \ref{dd-zz-xsec}a.
Since the processes $q \bar q \to Z^0Z^0$ is mediated, at lower order,
via t-channel
diagrams, it can be used to verify experimentally the $SM$ but cannot
serve as a probe for the search of s-channel massive gauge bosons
like the $Z'$ or
the extra dimension $Z^{\star}$  
which are dealt with in the following section.
As for the $gluon-gluon$ contribution to the $Z^0Z^0$ production in 
$pp$ collisions at 14 TeV,
it is estimated to be of the order of 15$\%$\cite{csc,ggpp}.\\ 

\begin{figure}
\centering{\psfig{figure=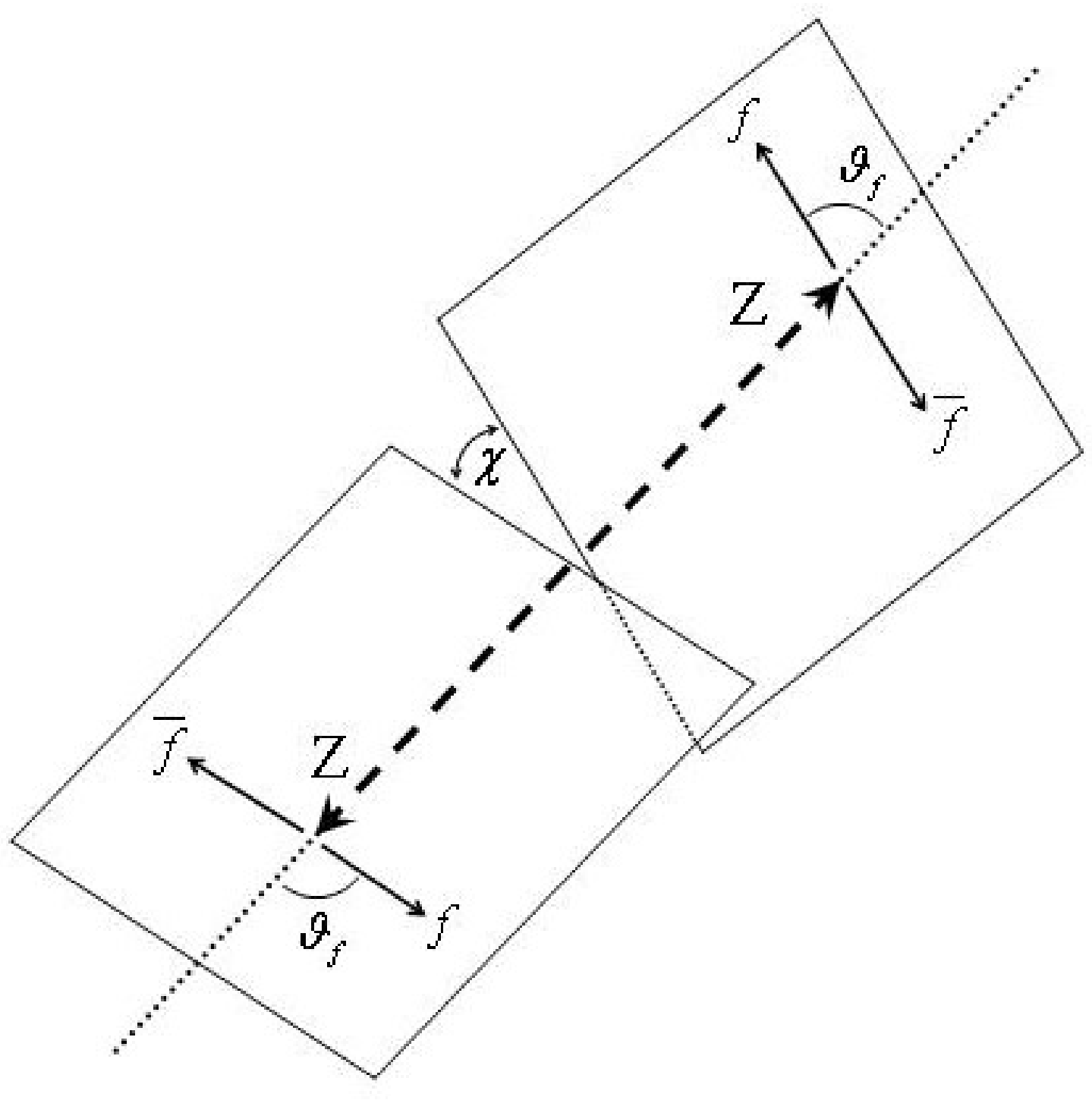,height=6.5cm} \ \ \ \ \
{\psfig{figure=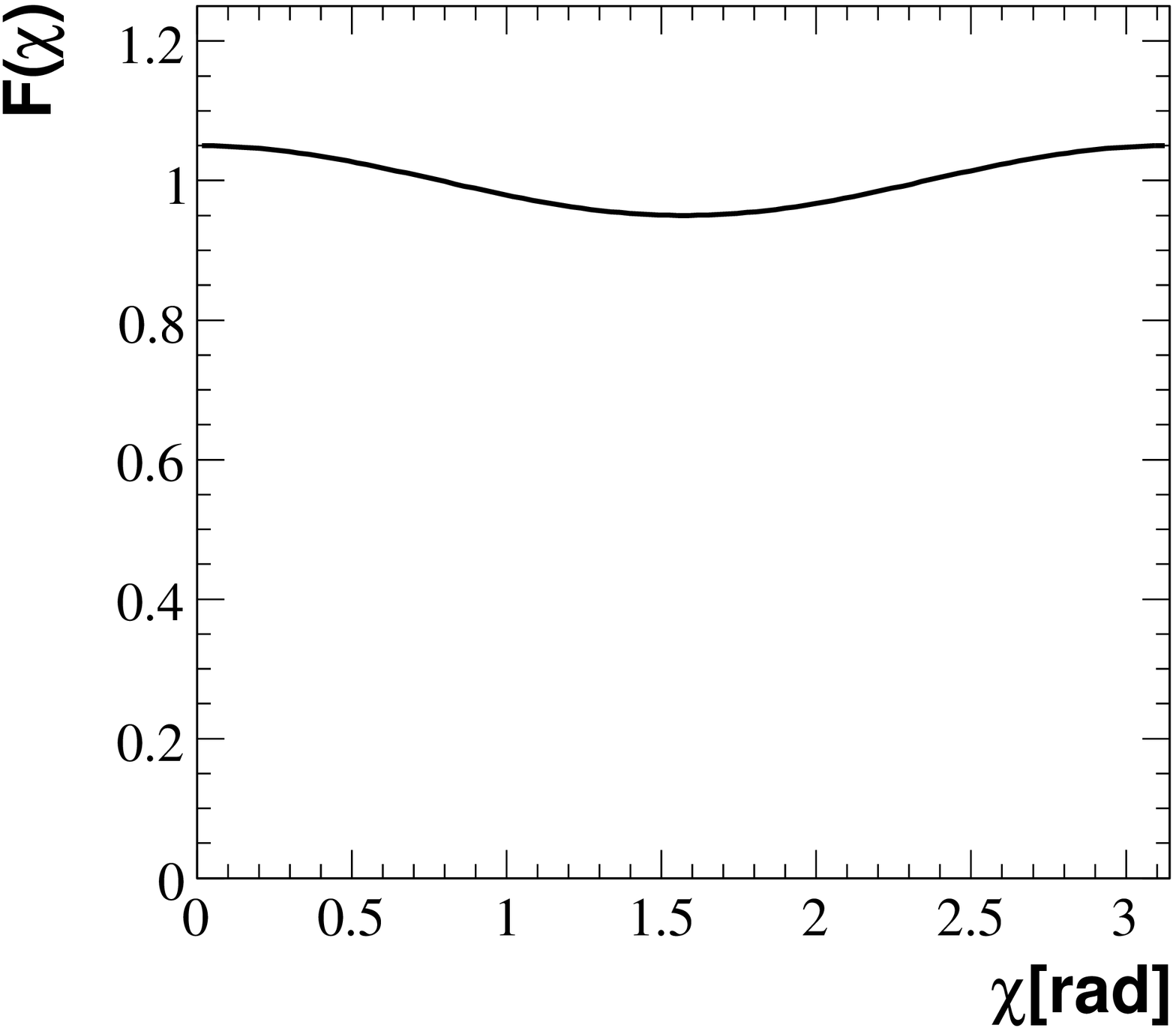,height=6.5cm}}} 
\caption{\small Left: The angle $\chi$ between the two $Z^0 \to f \bar
f$ decay
planes in the reaction $q \bar q \to Z^0Z^0$ (see text); Right: 
The $SM$ expectation of the angular distribution
$F(\chi)$ at $Ecm$=200 GeV of the $Z^0Z^0$ system, averaged over the production 
angle $cos\theta$.  
} 
\label{ee-zz-chi}
\end{figure}

The reliability of our derived $EDF$ expressions to transform the
parton anti-parton cross sections to the corresponding $pp$ 
reactions is demonstrated in Fig. \ref{dd-zz-xsec}b.
In this figure the data distribution  
of the reaction $pp \to Z^0Z^0$ at $\sqrt{S}$=14 TeV is obtained from
a Pythia 6.403 generated Monte Carlo sample of $\sim$38,000 
events utilizing the $PDF$ version CTEQ6L1 \cite{CTEQ6L1}. 
In the same figure $\sigma(q \bar q \to Z^0Z^0)$ is shown by the 
continuous line where it is normalized to the area under the MC generated 
sample of the $pp$ collisions.
This $q \bar q$ cross section
is then transformed via the $EDF$ to obtain the 
corresponding $pp \to q \bar q\to Z^0Z^0$ distribution (dotted line) 
which again is normalized to area defined by the MC data.
To be seen, there is an over all
agreement between the MC sample and the expected
$pp \to Z^0Z^0$ distribution. 
The slight deviation between the 
$EDF$ treated cross section and the
MC generated sample distribution 
can be traced among other reasons to the approximate numerical
evaluation of the integrals given in Eq. \ref{eq_p} and to  
the difference between
the $PDF$ version incorporated in the MC program and that 
used by us.\\ 

The $Z^0Z^0$ single and 
joint longitudinal polarizations, which are shown in
Fig. \ref{dd-zz-xsec}c, decrease with energy and are
seen to approach each other as they reach a nearly zero value
at 750 GeV. Next  
we calculate the single $Z^0$  
longitudinal polarization using 
the generated MC sample applying the known
helicity  projection operator $\Lambda_{00}$ \cite{opal_w_triple}.
The results modified by the $EDF$ are shown in Fig. \ref{dd-zz-xsec}d for five 
$Ecm$ regions, each having about 900 events, 
where they are seen to follow the general behavior of the
$SM$ expectation shown by the line drawn in the figure.\\
 
The reaction  $q \bar q \to Z^0Z^0$, which finally decays into
four charged ($e^{\pm}, \mu^{\pm}$) leptons, offers another 
measurable variable  
namely, the
distributions of the angle $\chi$ between the two decay planes
as illustrated in Fig. \ref{ee-zz-chi}. These planes are obtained
by first transforming the whole event configuration to the 
$Z^0Z^0$ center of mass system and then transforming each lepton decay pair
to the center of mass of its parent gauge boson. 
The angular distribution
$F(\chi)$, which is invariant under the transformation
$\chi \to \pi -\chi$, has the form \cite{kane}
\begin{equation}
F(\chi)\ =\ 1 + D cos(2\chi)\ ,
\end{equation}
where
\begin{equation}
D\ =\ 0.25 \times(\rho_{++--} +\rho_{--++})\ .
\end{equation}
As seen  from Fig. \ref{ee-zz-chi}, in the framework of the $SM$
the $F(\chi)$ distribution at $Ecm$=200 GeV, averaged over the
production angle, deviates
only slightly from uniformity and thus should be very
hard to detect. The study of $F(\chi)$ however may be quite
useful in the case where the $Z^0Z^0$ pair is the decay product
of a Higgs boson \cite{higgs}. 

\section{Detection of massive like$-$gauge vector bosons}
\label{massive}
In this section we discuss the expected effects due to the presence of
a massive vector boson
in the s-channel on the produced $SM$ di-gauge bosons final states. 
For the search of directly produced  
beyond the $SM$ gauge$-$like massive vector bosons, the reader
is advised to consult e.g. Ref. \cite{rizzo} and its references therein. 
 
\subsection{ The $Z'$ boson effect on the $W^+W^-$ final state} 
\label{zprime}
The possible existence of heavy $Z'$ vector bosons have been,
and still are, speculated in the framework  
of several models \cite{altarelli,zprime}. Searches for such heavy bosons
were carried out in LEP2 and in the Tevatron which reported 
a lower mass limit in the region between 0.8 to 1 TeV
\cite{langacker}. Whereas the coupling of a $Z'$ to 
fermions is expected to be similar to that of the Standard Model
$Z^0$  
($Z^0_{SM}$), its coupling to $W^+W^-$ is estimated to be much
weaker, namely by a factor of about 100 to 1000 \cite{zprime,langacker}.\\ 

\begin{figure}
\centering{\psfig{figure=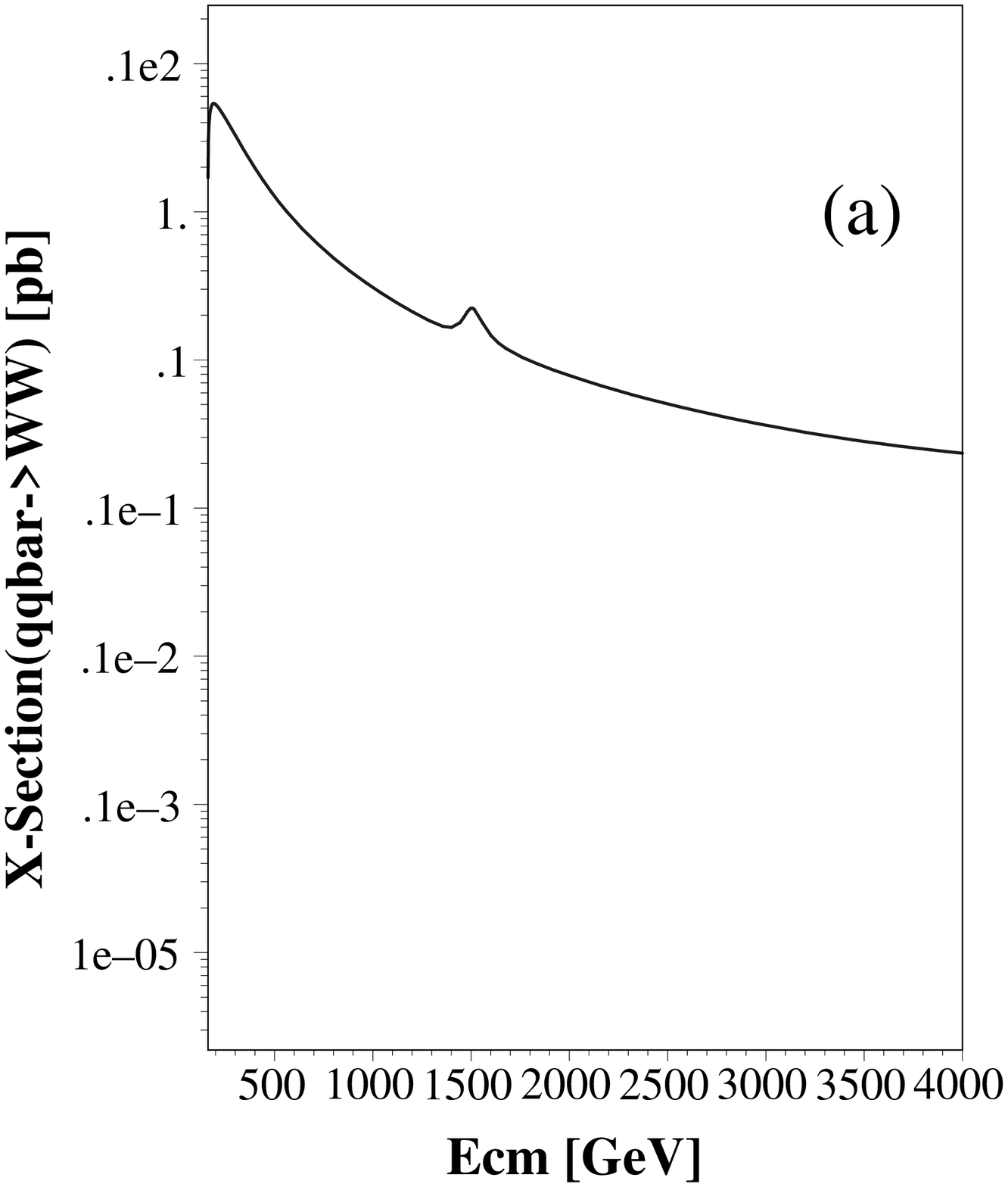,height=6.5cm,angle=0}\ \ \   
\psfig{figure=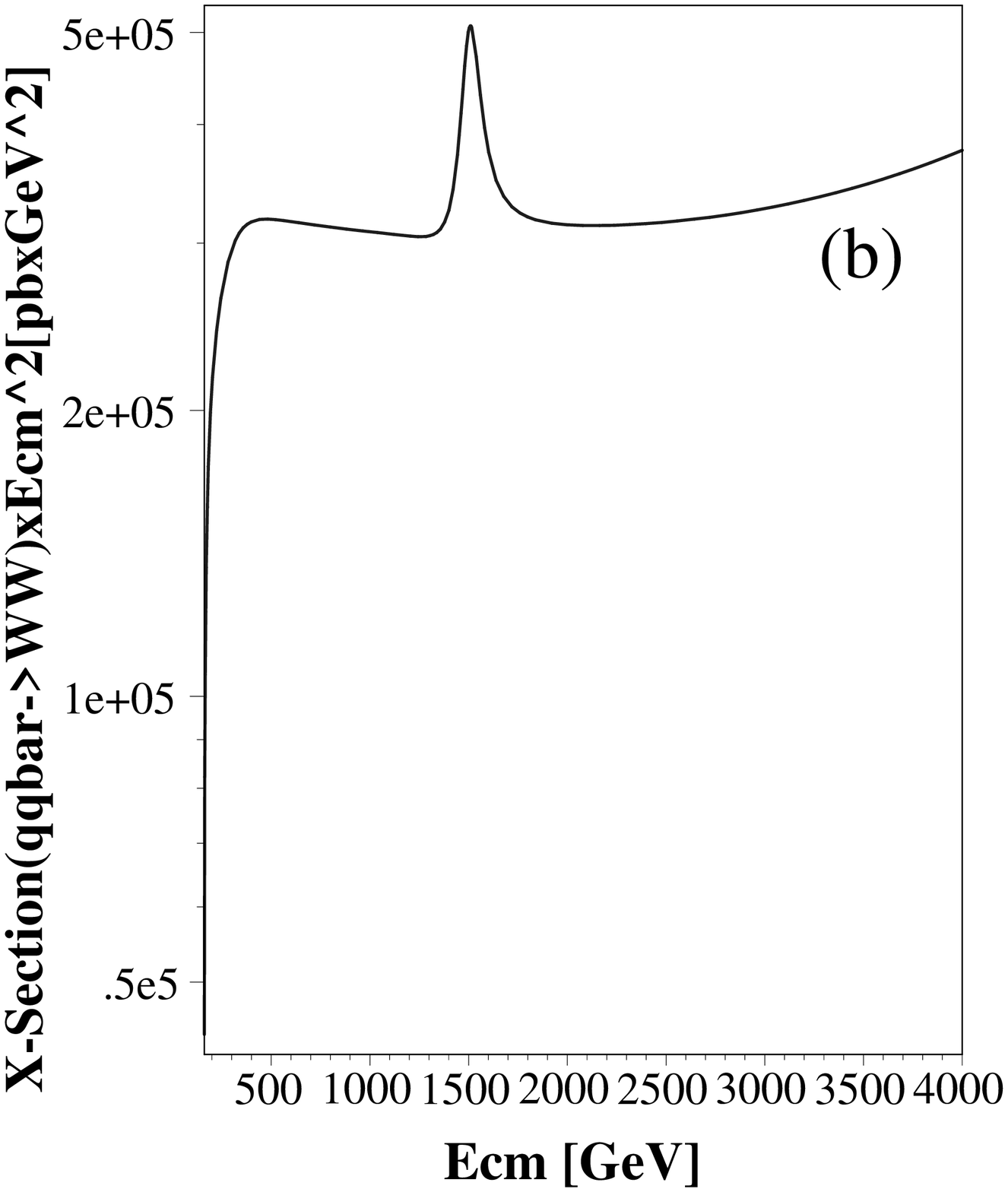,height=6.5cm,angle=0}\\  
\vspace{2mm}
\psfig{figure=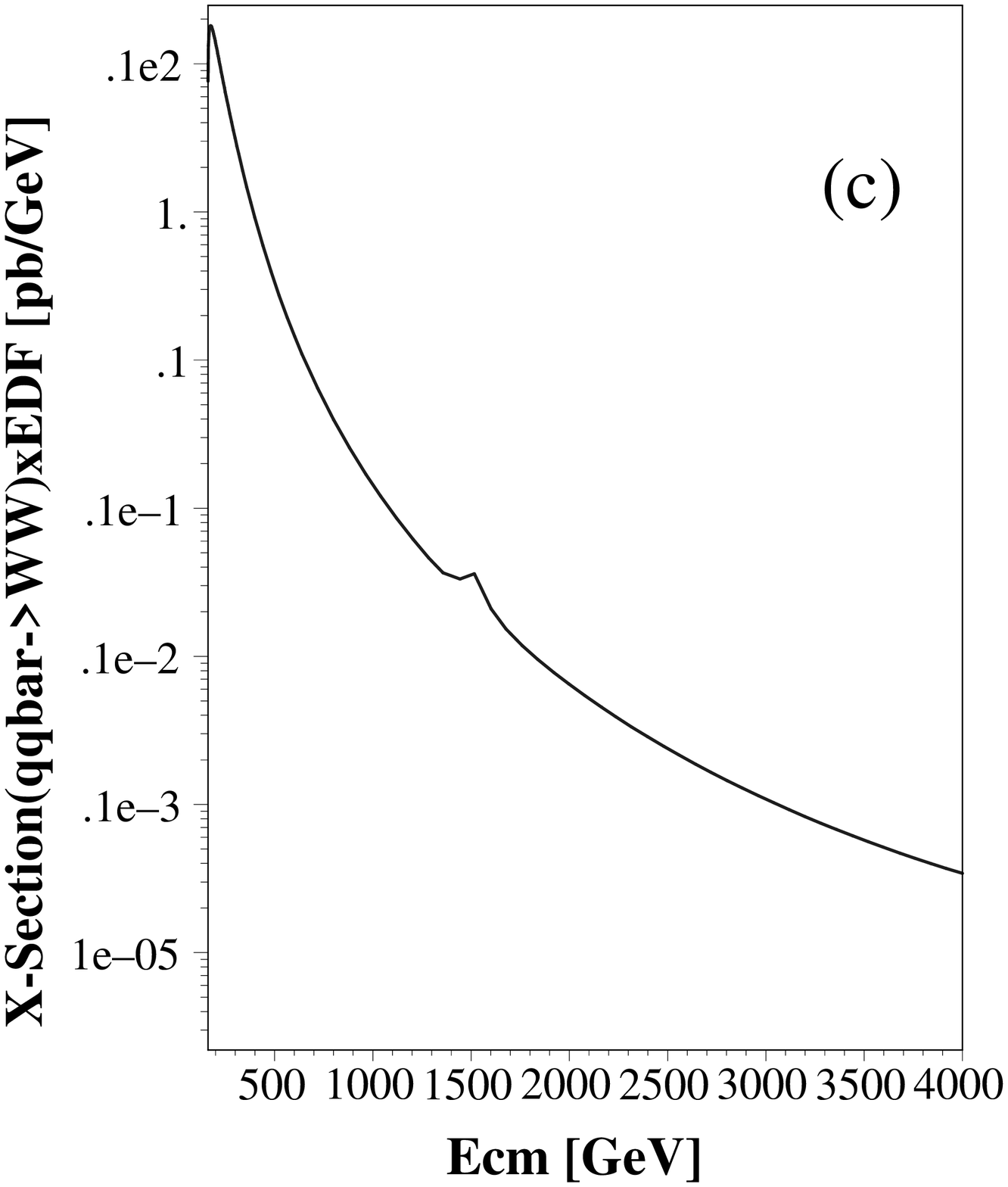,height=6.5cm,angle=0}\ \ \
\psfig{figure=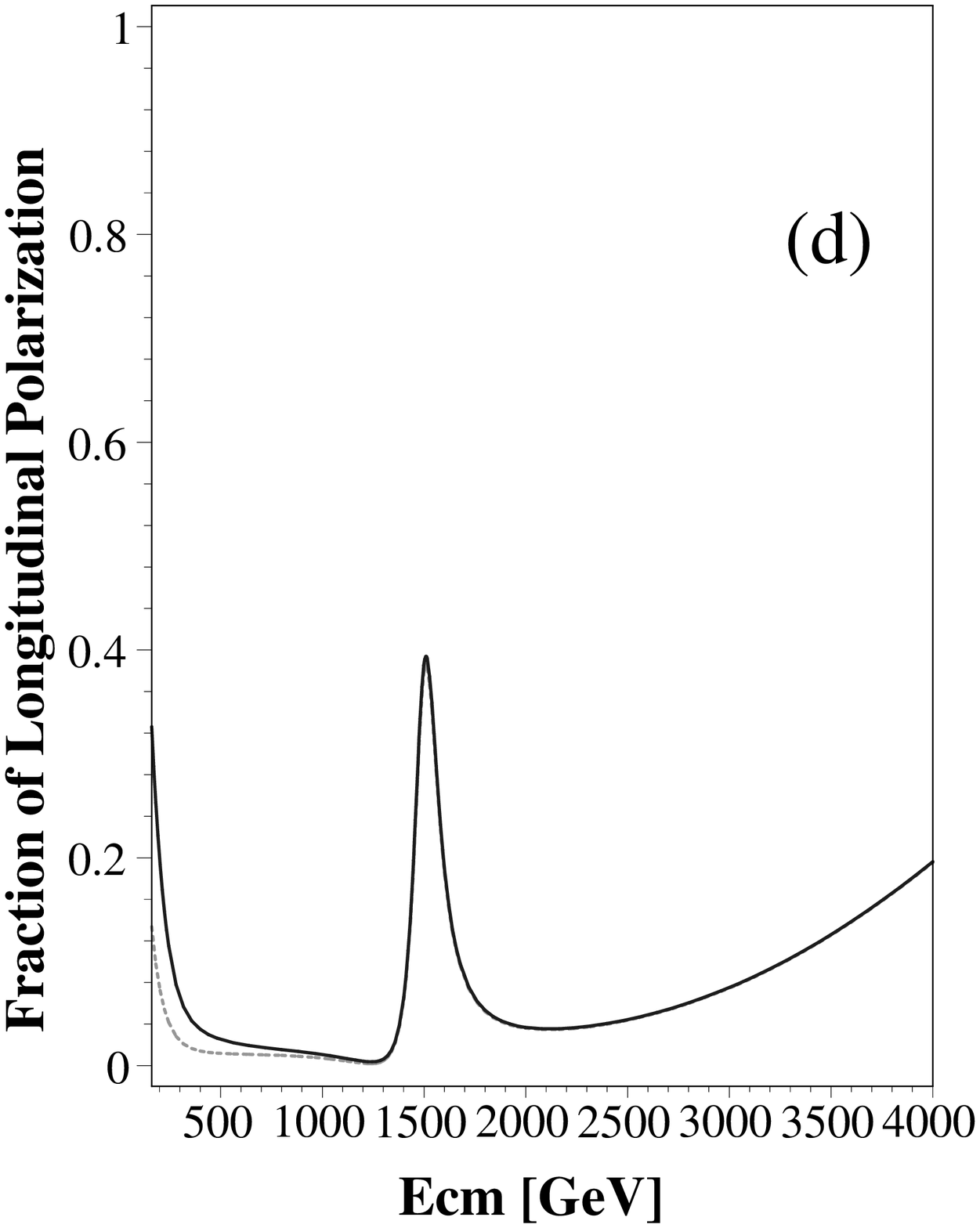,height=6.5cm}} 
\caption{\small Features of the 
reaction $\bar q q \to W^+W^-$ in the presence of an s-channel
$Z'$, of a  1.5 TeV mass and a 
width of 120 GeV, as a function of the $WW$ $Ecm$. 
(a): $\sigma(\bar q q \to W^+W^-)$;
(b): As (a) but multiplied by $Ecm^2$ to check unitarity;
(c): The corresponding $\sigma(pp \to W^+W^-)$ at $\sqrt{S_{pp}}$=14
TeV modified by $EDF$ from $\sigma(q \bar q \to W^+W^-)$;
(d): The single (continuous line) and
joint (dashed line) longitudinal polarizations as a function of
the $WW$ $Ecm$.
}
\label{uni_prime}
\end{figure}
To illustrate the feasibility  to detect a $Z'$ boson in the
reaction $pp \to \gamma /Z_{SM}^0 /Z' \to W^+W^-$ we attributed to the $Z'$
a mass of 1.5 TeV with a width of 120 GeV having a $SM$ like coupling to fermions 
while setting its
coupling to the $W^+W^-$ pair weaker by a factor of 500 than that
of the $Z_{SM}^0$ boson. In 
Fig. \ref{uni_prime}a   
$\sigma(\bar q q \to \gamma/Z_{SM}^0/Z' \to W^+W^-)$ is shown and
in Fig. \ref{uni_prime}b  
$\sigma(\bar q q \to\gamma/ Z_{SM}^0/Z' \to W^+W^-)\times Ecm^2$ is
presented to check the unitarity requirement \cite{uni_test}. 
The corresponding distribution of
$\sigma(pp \to \gamma/Z_{SM}^0/Z' \to W^+W^-)$ arising from
$pp$ collisions at the LHC is given in Fig. \ref{uni_prime}c. 
From this figure it is clear
that the cross section signal arising from the presence of
a $Z'(1.5)$ in the s-channel is by far too weak to be experimentally
detected if only for the reason that the $pp$ luminosity 
at the LHC will
be known to $\approx 10\%$. On the other hand, 
the longitudinal polarization of the
$W^+W^-$ final state, shown in Fig. \ref{uni_prime}d, 
reveals a signal that is by approximately a factor ten larger than the $SM$
one as is estimated from Fig. \ref{dubar-wz-xsec}c. Furthermore, the
longitudinal polarization is seen to increase with $Ecm$
beyond the $Z'(1.5)$ energy unlike the case where only the $Z^0_{SM}$ 
contributes in the s-channel to the $W^+W^-$ final state.\\ 

In addition to the possible existence of the $Z'$, 
the conjecture of
Kaluza Klein Extra Dimensions does also envisage the existence of 
massive excited $Z$ vector bosons, labeled as $Z^{\star}$. 
These couple to fermions as 
the $Z_{SM}^0$ \cite{extra} and the mass of the 
lightest one of them is currently estimated to be equal or larger than
$\sim$4 TeV. 
Inasmuch that these massive vector bosons
do have a sufficiently strong coupling to the $W^+ W^-$ boson pair, 
they may be detected as a
resonance in the $W^+W^-$ invariant mass distributions. If their
coupling strength to $W^+W^-$ 
is similar to that chosen by us for the $Z'$, then their 
longitudinal polarization behaviour follows 
the one shown in Fig. \ref{uni_prime}d and could be instrumental in the
detection of these  massive vector bosons.

\subsection{ The $W'$ boson effect on the $W^{\pm} Z^0$ final states}
\label{wprime}
The existence of massive $W'$ or a $W'-$like bosons are now a 
common prediction of
several beyond the $SM$ physics scenarios where their properties, 
and in particular their couplings to fermions and their
trilinear coupling $W'WZ$, are dealt with in various
investigations \cite{rizzo}.\\ 
\begin{figure}
\centering{\psfig{figure=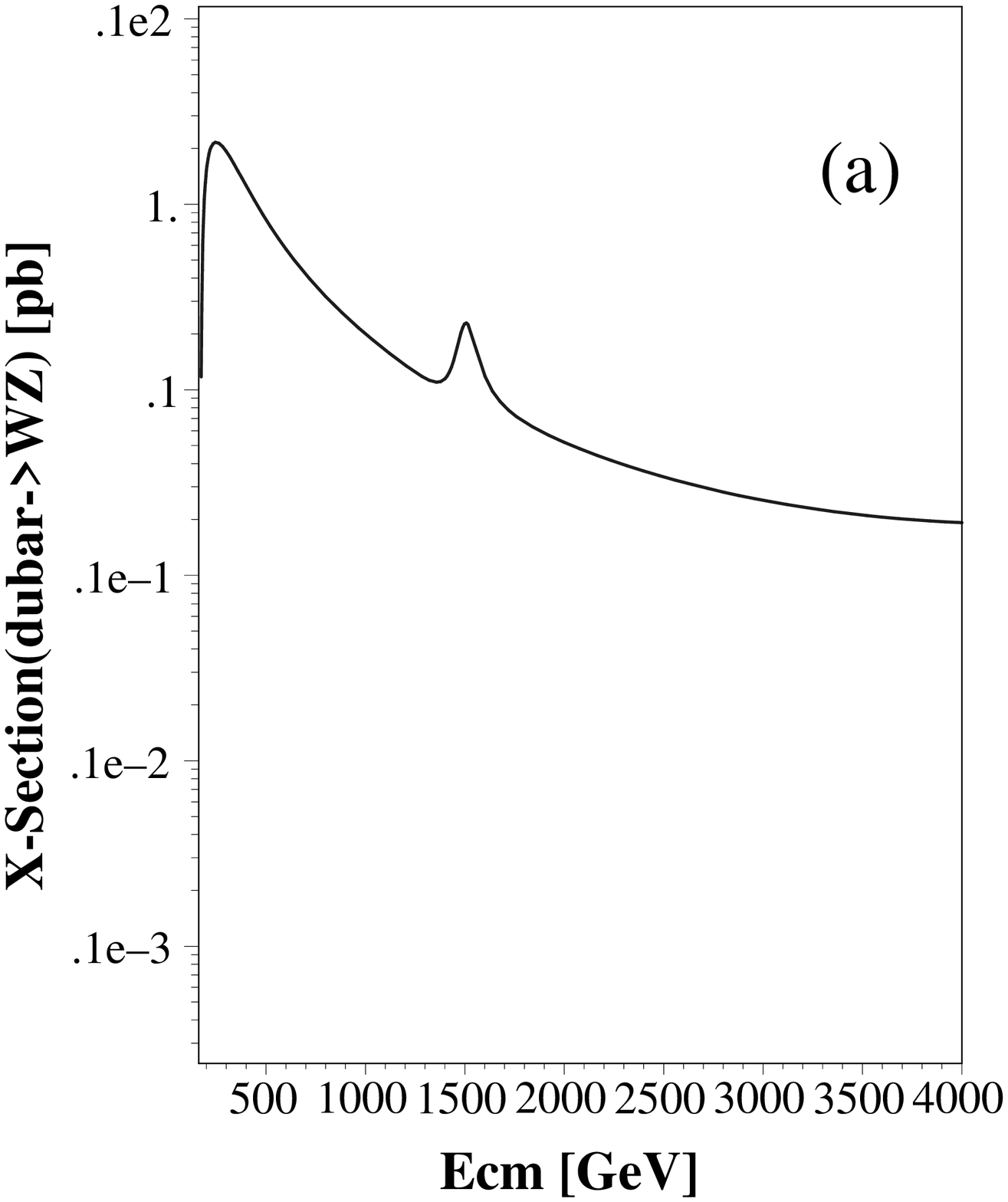,height=7cm} \ \ \  
\psfig{figure=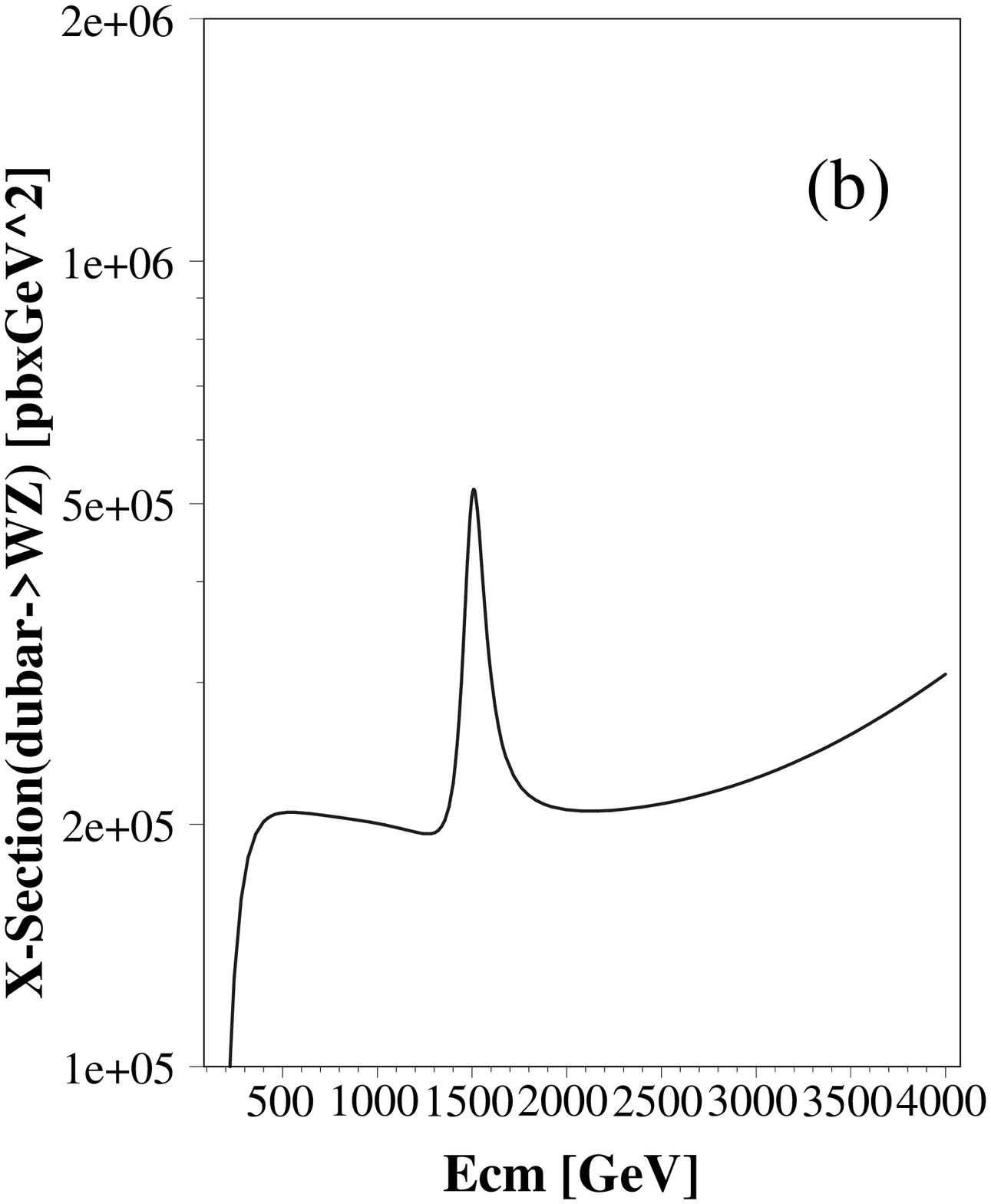,height=7cm}\\ 
\vspace{2mm}
\psfig{figure=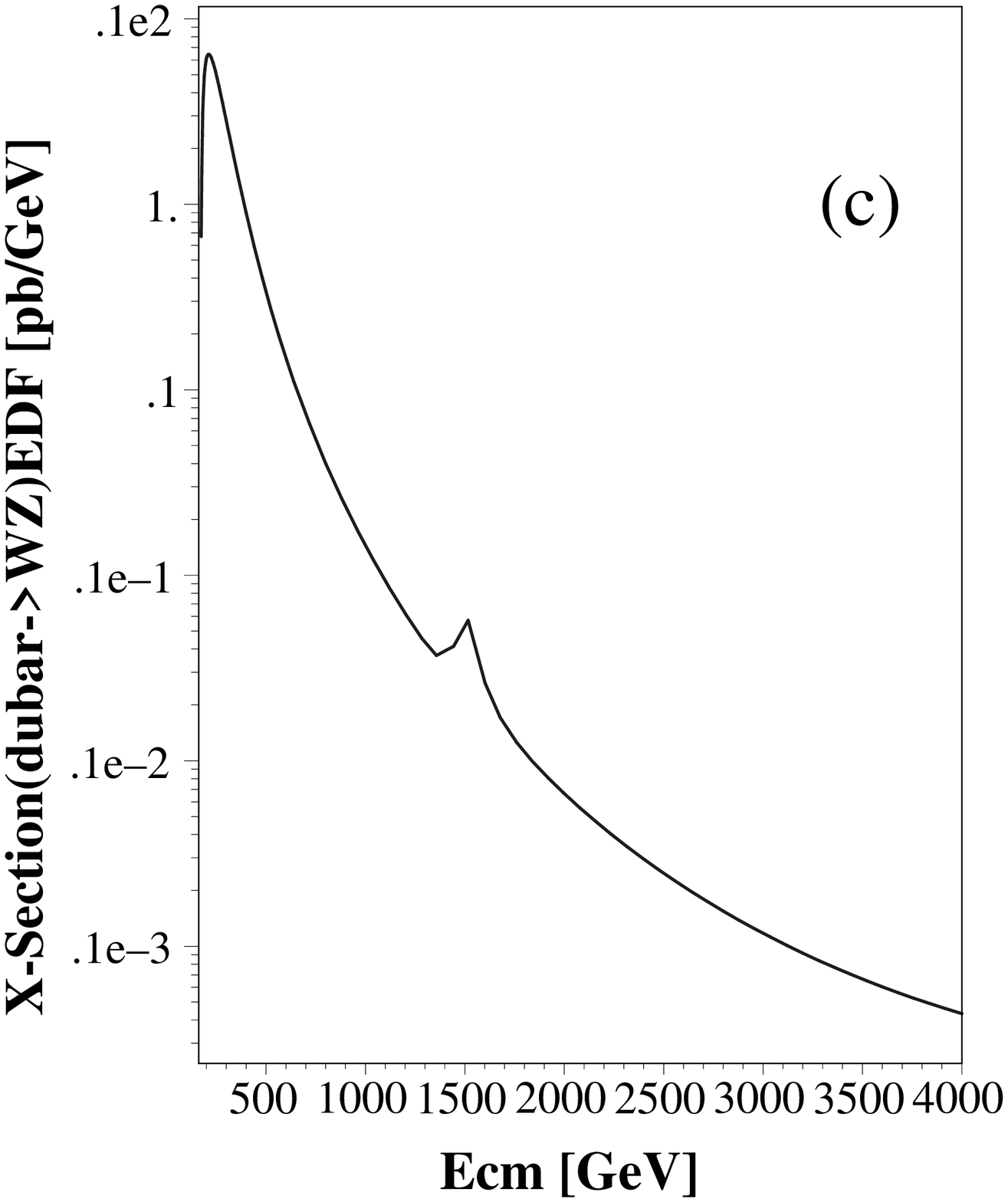,height=7cm} \ \ \
{\psfig{figure=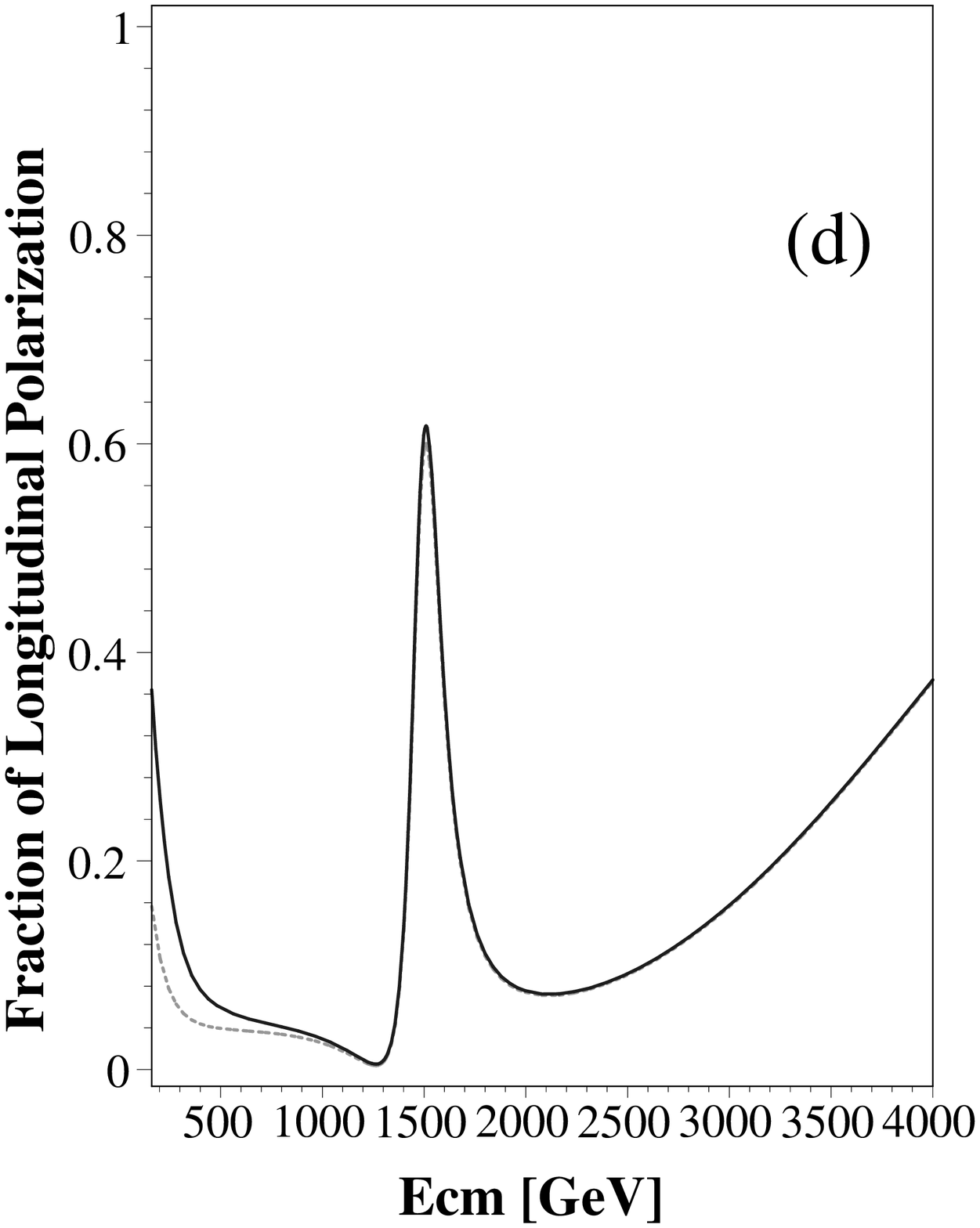,height=7cm}}}
\caption{\small Features of the 
reaction $d\bar u \to W^-Z^0$ in the presence of
a 1.5 TeV $W'$ with a width of 120 GeV as a function of the
$WZ$ $Ecm$.
(a): $\sigma(d\bar u \to W^-Z^0)$;
(b): $\sigma(d\bar u \to W^-Z^0)\times Ecm^2$ for unitarity check; 
(c): $\sigma(pp \to W^-Z^0)$ at $\sqrt{S_{pp}}$=14 TeV modified via 
$EDF$ from 
$\sigma(d\bar u \to W^-Z^0)$;
(d): The longitudinal polarization.
The solid line is  $\rho_{00}$ the single $Z$ polarization,  
and the
dashed line is $\rho_{0000}$, the joint $W^-Z^0$ polarization.    
} 
\label{dubar-wz-wp}
\end{figure}\\
Here we examine the effect of the presence  
of a 1.5 TeV $W'$ vector boson in the s-channel 
having a 120 GeV width which has a $SM$ like
couplings to fermions but a $W'WZ$ coupling smaller by a factor
500 than the corresponding $WWZ$ one. The resulting
cross section for the reaction $d \bar u \to W/W' \to W^-Z^0$
is shown in Fig. \ref{dubar-wz-wp}a.  
This cross section, multiplied by $Ecm^2$ is shown in  Fig. \ref{dubar-wz-wp}b
to verify the unitarity condition. The signal of 
$W'$ presence in the $W^-Z^0$ production 
in the $pp$ reactions, seen in Fig. \ref{dubar-wz-wp}c, is far too weak
to be noticed. On the other hand, the longitudinal polarization
signal at 1.5 TeV and its behaviour
at energies above it (see Fig. \ref{dubar-wz-wp}d) is considerably higher than that expected in
the absence of a $W'$ gauge boson in the s-channel.   

\section{Summary}
\label{summary}
In the frame work of the Standard Model the reactions
$q \bar q \to VV'$ are calculated via the helicity amplitudes for
the final states $W^+W^-$, $W^{\pm}Z^0$ and $Z^0Z^0$ to obtain
their cross sections and longitudinal polarizations as a function
of their center of mass energy, $Ecm$. These cross sections are
transformed to the corresponding ones expected to be observed in
$pp$ collisions at 14 TeV, by using our parameterizations for 
the $q\bar q$ Energy Density
Functions. These 
expressions for the $EDF$ may also be useful for the opposite
transformation, that is, from the $pp$ processes to the basic parton
anti-parton cross sections.\\

Whereas in the LHC the cross section measurement accuracy  
depends above all on the luminosity precision, currently
estimated to be $10\%$,   
clearly one of the polarization measurement virtues 
is its  
independence of the luminosity.  
The single and joint longitudinal polarization of the $VV'$ final
states in $pp$ collisions are seen to 
decrease and approach each other as $Ecm$ increases. Whereas
at present the polarization measurements of the $W^+W^-$ final state 
are not feasible, they are accessible in the $Z^0Z^0$ channel. 
Methods to measure to a good approximation the longitudinal
polarizations of the $W^{\pm}Z^0$ final state is likely to be worked
out in the future.\\

The effect on the production of the $VV'$ states
from the existence of s-channel massive bosons like the
$Z'$, $W'$ and the $ED$ heavy $Z^{\star}$, is
studied with the results that the change in the polarization structures 
are more pronounced than those seen in the behaviour of the cross sections.\\ 

\noindent
{\bf Acknowledgements}\\   
Our thanks are due to members of the Tel-Aviv University ATLAS group,
and in particular to E. Etzion, for their continuous support throughout this
work. In addition we are grateful to Y. Oz and S. Nussinov
for their helpful discussions and suggestions.   
%\vspace{3cm}
%\newpage        

\end{document}